\documentclass{vldb}

\usepackage{graphicx}
\usepackage{balance}  

\usepackage{longtable}
\usepackage[latin1]{inputenc}
\usepackage{latexsym}

\usepackage{amsmath, amsthm, amssymb}
\usepackage{amsfonts}
\usepackage[format=plain,indention=.6cm, font=small, labelfont=bf,belowskip=-10pt,aboveskip=0pt]{caption}
\usepackage[labelformat=simple,skip=0pt]{subcaption}
\usepackage{fancyhdr}

\usepackage{lscape}
\usepackage[T1]{fontenc}
\usepackage{algcompatible}
\usepackage{floatrow}
\usepackage{algorithm}
\usepackage{multirow}
\usepackage{wrapfig}
\usepackage{url}

\newsavebox{\boxone}
\newsavebox{\boxtwo}
\newsavebox{\boxthree}

\newlength{\narrow}
\newlength{\cnarrow}
\newcommand{\topline}{
  \hrule
  \vskip .5\baselineskip}
\newcommand{\bottomline}{
  \vskip 2pt
  \hrule}

\newcommand{\chbox}[2]{
  \hbox to #1{\hss\vtop{#2}\hss}}

\newcommand{\nchbox}[1]{
  \chbox{\narrow}{#1}}

\newcommand{\cnchbox}[1]{
  \chbox{\cnarrow}{#1}}

\newcommand{\fcode}[1]{
  
  \chbox{\textwidth}{\tgrind\input{#1}}}

\newcommand{\ncode}[1]{
  
  \chbox{\narrow}{\tgrind\input{#1}}}

\def\nfig#1#2#3{
  \vtop{\nchbox{#1}
  \hbox to\narrow{\parbox{\narrow}{\caption{#2}\label{#3}}}}}

\newcommand{\cncode}[1]{
  \chbox{\cnarrow}{\tgrind\input{#1}}}

\def\codefiggen[#1]#2#3#4#5#6{
  \begin{figure}[#1]
  #5
  \fcode{#2}
  \center\parbox{.9\textwidth}{\caption{#3}\label{#4}}
  #6
  \end{figure}}

\def\codefig[#1]#2#3#4{
  \codefiggen[#1]{#2}{#3}{#4}{}{}}

\def\codefigline[#1]#2#3#4{
  \codefiggen[#1]{#2}{#3}{#4}{\topline}{\bottomline}}

\def\doublefiggen[#1]#2#3#4#5#6#7#8#9{
  \begin{figure}[#1]
  #8
  \hbox to \textwidth{
  \nfig{#2}{#3}{#4}
  \hfil
  \nfig{#5}{#6}{#7}}
  #9
  \end{figure}}

\def\doublefig[#1]#2#3#4#5#6#7{
  \doublefiggen[#1]{#2}{#3}{#4}{#5}{#6}{#7}{}{}}

\def\doublefigline[#1]#2#3#4#5#6#7{
  \doublefiggen[#1]{#2}{#3}{#4}{#5}{#6}{#7}{\topline}{\bottomline}}

\def\doublecodefig[#1]#2#3#4#5#6#7{
  \doublefig[#1]{\ncode{#2}}{#3}{#4}{\ncode{#5}}{#6}{#7}}

\def\doublecodefigline[#1]#2#3#4#5#6#7{
  \doublefigline[#1]{\ncode{#2}}{#3}{#4}{\ncode{#5}}{#6}{#7}}

\newcommand{\codepair}[4]{\vbox{
  \hbox{\ncode{#1} \hfil \ncode{#3}}
  \vskip .3\baselineskip plus .3\baselineskip
  \hbox{\hbox to\narrow{#2\hfil} \hfil \hbox to\narrow{#4\hfil}}}}

\def\codepairfig[#1]#2#3#4#5#6#7{
  \begin{figure}[#1]
  \codepair{#2}{#3}{#4}{#5}
  \center\parbox{.9\textwidth}{\caption{#6}}
  \label{#7}
  \end{figure}}

\def\cncodepairfiggen[#1]#2#3#4#5#6#7{
  \begin{figure}[#1]
  #6
  \hbox{\cncode{#2}\hfil\cncode{#3}}
  \center\parbox{.9\columnwidth}{\caption{#4}\label{#5}}
  #7
  \end{figure}}

\def\cncodepairfig[#1]#2#3#4#5{
  \cncodepairfiggen[#1]{#2}{#3}{#4}{#5}{}{}}

\def\cncodepairfigline[#1]#2#3#4#5{
  \cncodepairfiggen[#1]{#2}{#3}{#4}{#5}{\topline}{\bottomline}}

\def\doublefigOnecap*[#1]#2#3#4#5{
  \begin{figure*}[#1]
  \hbox to \textwidth{
  \nchbox{#2}
  \hfil
  \nchbox{#3}}
  \caption{#4}
  \label{#5}
  \end{figure*}}

\def\doublefigOnecap[#1]#2#3#4#5{
  \begin{figure}[#1]
  \topline
  \hbox to \columnwidth{
  \cnchbox{#2}
  \hfil
  \cnchbox{#3}}
  \caption{#4}
  \label{#5}
  \bottomline
  \end{figure}}

\def\PSfig[#1]#2#3#4{
 \begin{figure}
 \centerline{\psfig{file=#2,width=\columnwidth}}
 \caption{{#3}}
 \label{#4}
 \end{figure}}

\def\PSfiglines[#1]#2#3#4{
 \begin{figure}[#1]
 \topline
 \centerline{\psfig{file=#2,width=\columnwidth}}
 \caption{{#3}}
 \label{#4}
 \bottomline
 \end{figure}}

\def\PSfiglinesht[#1]#2#3#4#5{
 \begin{figure}[#1]
 \topline
 \centerline{\psfig{file=#2,height=#3}}
 \caption{{#4}}
 \label{#5}
 \bottomline
 \end{figure}}

\def\doublePSfig[#1]#2#3#4#5#6{
  \doublefigOnecap[#1]
    {\cnchbox{\psfig{file=#2,height=#4}}}
    {\cnchbox{\psfig{file=#3,height=#4}}}
    {#5}
    {#6}}

\newlength{\boxwidth}
\newcommand{\bproof}{{\bf Proof Sketch: }}
\newcommand{\eproof}{\mbox{$\Box$}}

\def\tabdoublecode#1#2#3#4{
 \begin{figure*}[t]
 \topline\vs{-.4}
 \hbox to \columnwidth{
 \vtop{\small
 \begin{tabbing}
 #1
 \end{tabbing}}
 \hfil
 \hfil
 \hfil
 \vtop{\small
 \begin{tabbing}
 #2
 \end{tabbing}}
 }
 \caption{#3\label{#4}}
 \bottomline
 \end{figure*}
}
\def\tabtriplecode#1#2#3#4#5{
 \begin{figure}
 \topline\vs{-.4}
 \hbox to \columnwidth{
 \vtop{\small
 \begin{tabbing}
 #1
 \end{tabbing}}
 \hfil
 \hfil
 \hfil
 \vtop{\small
 \begin{tabbing}
 #2
 \end{tabbing}}
 \hfil
 \hfil
 \hfil
 \vtop{\small
 \begin{tabbing}
 #3
 \end{tabbing}}
 }
 \caption{#4\label{#5}}
 \bottomline
 \end{figure}
}

\newtheorem{lemma}{Lemma}
\newcommand{\blemma}{\begin{lemma}}
\newcommand{\elemma}{\end{lemma}}

\newtheorem{thm}{Theorem}
\newcommand{\bthm}{\begin{thm}}
\newcommand{\ethm}{\end{thm}}

\newtheorem{defin}{Definition}
\newcommand{\bdefin}{\begin{defin}}
\newcommand{\edefin}{\end{defin}}

\newtheorem{observation}{Observation}
\newcommand{\bobserv}{\begin{observation}}
\newcommand{\eobserv}{\end{observation}}

\newtheorem{coroll}{Corollary}
\newcommand{\bcoroll}{\begin{coroll}}
\newcommand{\ecoroll}{\end{coroll}}

\newcommand{\vs}[1]{\vspace{#1cm}}
\newcommand{\be}{\begin{equation}}
\newcommand{\ee}{\end{equation}}

\newcommand{\bdesc}{\begin{description}}
\newcommand{\edesc}{\end{description}}
\newcommand{\benum}{\begin{enumerate}}
\newcommand{\eenum}{\end{enumerate}}
\newcommand{\bitem}{\begin{itemize}}
\newcommand{\eitem}{\end{itemize}}
\newcommand{\bcenter}{\begin{center}}
\newcommand{\ecenter}{\end{center}}
\newcommand{\btabular}{\begin{tabular}}
\newcommand{\etabular}{\end{tabular}}
\newcommand{\beqnarr}{
 \begin{eqnarray}}
\newcommand{\eeqnarr}{\end{eqnarray}}

\begin{document}



\title{Pruned Landmark Labeling Meets Vertex Centric Computation: A Surprisingly Happy Marriage!}

\numberofauthors{1}
\author{\small \alignauthor Ruoming Jin$^\dagger$~~Zhen Peng$^\star$~~Wendell Wu$^\dagger$~~Feodor Dragan$^\dagger$~~Gagan Agrawal$^\ddagger$~~Bin Ren$^\star$ \\
\affaddr{$^\dagger$\mbox{ }Kent State University ~~~~~~~~~~~~ $^\star$\mbox{ }College of William and Mary  ~~~~~~~~~ $^\ddagger$\mbox{ }The Ohio State University } \\
\email{ \{rjin1,wwu12,fdragan\}@kent.edu~
\{zpeng,bren\}@cs.wm.edu~
agrawal.28@osu.edu}
}

\maketitle
\begin{abstract}
In this paper, we study how the Pruned Landmark Labeling (PPL) algorithm can be parallelized in a scalable 
fashion, producing the same results as the sequential algorithm. More specifically, we parallelize 
using a Vertex-Centric (VC) computational model on  a modern SIMD powered multi-core architecture. We design a new VC-PLL algorithm that resolves the apparent mismatch between the inherent sequential dependence of the PLL algorithm and the Vertex-Centric (VC) computing model. Furthermore, we introduce a novel batch execution model for VC computation and the BVC-PLL algorithm to reduce the computational inefficiency in VC-PLL. Quite surprisingly, the theoretical analysis reveals that under a 
reasonable  assumption, BVC-PLL has lower computational and memory access costs than PLL and indicates it may run faster than PLL as a sequential algorithm. We also demonstrate how BVC-PLL algorithm can be extended to handle directed graphs and weighted graphs and how it can utilize the hierarchical parallelism on a modern parallel computing architecture. Extensive experiments on real-world graphs not only show the sequential BVC-PLL can run more than two times faster than the original PLL, but also demonstrates its parallel efficiency and scalability. 
\end{abstract}

\section{Introduction}\label{sec:intro}

Computing the shortest path distance between any two vertices stands out as one of the most fundamental graph operators in querying and analyzing massive graphs, with applications ranging from transportation systems, social networks, software systems, the WWW,  and semantic web,  
among others. This operation also serves as the basis for more complex graph analytics and mining operations, such as graph pattern matching~\cite{DBLP:journals/tkde/ChengYY11,Zou:2009:DPM}, distance join processing~\cite{Sankaranarayanan:2006:DJQ}, and centrality computation~\cite{Brandes01afaster}.  

Distance computation on road networks has become a common service for internet map applications, such as Google Maps. However, computing shortest path distances over scale-free complex networks---for 
example, massive social and web graphs---remains  a challenging problem~\cite{akiba2013fast,DBLP:journals/corr/abs-1305-0507,Ouyang:2018:HME:3183713.3196913}. To provide the exact distance query result, the $2$-hop labeling approach~\cite{cohen2hop} has emerged as a major tool. Given a graph, it aims to assign each vertex $v$ a label $L(v)$  
comprising  a list of vertices and their distance to $v$. 
Subsequently,  given any two vertices, we can only use the label information, $L(u)$ and $L(v)$,  to recover their exact distance. 

Since the seminar work by Cohen~\cite{cohen2hop}, numerous efforts over a ten-year period   ~\cite{hopiedbt,ChengYLWY06,DBLP:conf/sigmod/JinRXL12,DBLP:conf/edbt/ChengYLWY08,Abraham:2011:HLA,Geisberger:2008,Sankaranarayanan:2009} have largely failed in making $2$-hop labeling practical  on real-world graphs with millions of vertices and edges, until the discovery of {\em Pruned Landmark Labeling (PLL)}~\cite{akiba2013fast}. 
This new labeling approach adopts a fast greedy process to iteratively assign each vertex (one vertex at a time according to certain vertex order) the label of other vertices with respect to a distance check criterion: a vertex $u$ can be added into a vertex $v$'s label $L(v)$ if there are no other prior recorded vertex $h\in L(u)$, such that it can provide equal or shorter distance: $u \in L(v)$ if $d(u,v) < d(u,h)+ d(h,v)$ for all $h \in L(u)$ and $h \in L(v)$. Here $L(u)$ and $L(v)$ are the partial labels being constructed before the labeling of vertex $u$. 
Once the labeling process is done, the results are guaranteed to be {\em minimum} as no hops need or can be removed to recover all the pairwise distance or reachability information. 

In the past few years, a number of studies ~\cite{delling2014robust,li2017experimental} have further validated and confirmed the scalability of this approach. As a result,  $2$-hop  
has  gone from handling  small graphs with thousands of vertices and edges to large graphs with millions of vertices and edges. 

\noindent{\bf PLL meets Vertex Centric Computation:}  
As modern computing architectures become increasingly parallel with more and more powerful GPUs and multicore with SIMD architectures emerging as over-the-shelf choices for graph processing/database systems and as the size of real-world graphs continue to grow bigger, an important question naturally arises: can PLL take full advantage of modern parallel computing architectures to better handle massive real-world graphs?  
Furthermore, since {\em vertex centric} (VC) computation               ~\cite{Malewicz:2010:PSL,mccune2015thinking} has become the de-facto 
standard for parallel graph processing and graph databases, can PLL be parallelized using the vertex centric scheme?  
As the industry adoption of graph databases and graph analytics systems is accelerating, answering these questions is becoming critical. 

However, the marriage between PLL and VC seems to be quite a mismatch: the original PLL algorithm is inherently sequential; i.e., the algorithm  
operates one vertex at a time to label the entire graph, and the labeling of a vertex depends on the partial labeling results from earlier processed vertices. Not only that, it has also been claimed  that PLL does not fit into a VC model~\cite{qiu2018parapll}. 

\noindent{\bf Parallel PLL:} Given the strong task dependency existing within a single vertex labeling process and across labeling vertices, all existing attempts on parallelization rely on the computational flow of the sequential PLL. 
The original PLL paper ~\cite{akiba2013fast} has suggested to simply parallelize the BFS labeling process of each vertex instead of dealing with inter-vertex labeling dependency.
Clearly, without the inter-vertex labeling parallelization, the parallelism is quite limited. The two recent attempts ~\cite{ferizovic2015parallel,qiu2018parapll} treat each vertex labeling process as a single task, and allow multiple vertices to simultaneously traverse and label other vertices sequentially  following the original PLL logic. Thus, the full benefits of PLL across the labeling order of vertices cannot be maintained and they cannot produce the same compact label as the original PLL.  


\noindent{\bf Our Contributions:}
In this paper, we study how PLL can be effectively parallelized (and more specifically,  under a Vertex-Centric (VC) computational model) using a modern SIMD powered multi-core architecture. 
Specifically, we study the following research problems and make several interesting discoveries along the way: 

\noindent{\bf 1. Parallel PLL Algorithm (Section~\ref{design-vc}):} To solve the mismatch between the inherent sequential/dependence of PLL algorithms and the VC model, we introduce a new VC-PLL algorithm that utilizes VC to parallelize PLL and is guaranteed to produce the same labels as original PLL. However, the performance of basic VC-PLL turns out to be quite disappointing compared to  PLL (both using a single thread). The theoretical analysis reveals two key factors in VC-PLL: message passing and remote memory access during vertex computation, which both introduce additional costs compared to the PLL algorithm.  

\noindent{\bf 2. Batched Vertex-Centric PLL(Section~\ref{sec:design-bvc}):} To deal with the limitations of VC-PLL, we introduce a novel batch execution model for vertex-centric computation and a new BVC-PLL algorithm which largely preserves the same vertex computation function while reducing the costs of message passing and remote memory access. Quite surprisingly, an in-depth
and apple-to-apple cost analysis between BVC-PLL and PLL reveals that under certain reasonable assumptions, {\em VC-PLL has lower computational costs and memory access costs than PLL}! This indicates BVC-PLL may run faster than PLL even as a sequential algorithm. 

\noindent{\bf 3. Generalization and System Optimization(Section~\ref{sec:system}):} We discuss how BVC-PLL can be extended to handle directed graphs and weighted graphs. We also study how BVC-PLL can be supported by modern parallel computing architecture using the {\em hierarchical parallelism}: the coarse grained
thread-level parallelism and the fine-grained data-level parallelism (i.e., SIMD parallelism or vectorization). 

\noindent{\bf 4. Experimental Study(Section~\ref{sec:eval}):} 
Our extensive evaluation focuses on the following questions: 
Can the BVC-PLL algorithm using only one thread (sequential execution) run faster than the original PLL? 
How does BVC-PLL scale to multiple threads  and how is  its parallel scalability compared to other parallel algorithms? 
What are the main factors affecting its performance? 
We show that the sequential BVC-PLL can run more than two times faster than the original PLL (both using one single thread)! Additionally, BVC-PLL also has good scalability and obtains close to linear speedup using $20$ threads on several real-world datasets.

\section{Preliminaries}\label{sec:bg}

\begin{figure*}[th]
\begin{subfigure}{0.32\textwidth}
    \includegraphics[width=\linewidth]{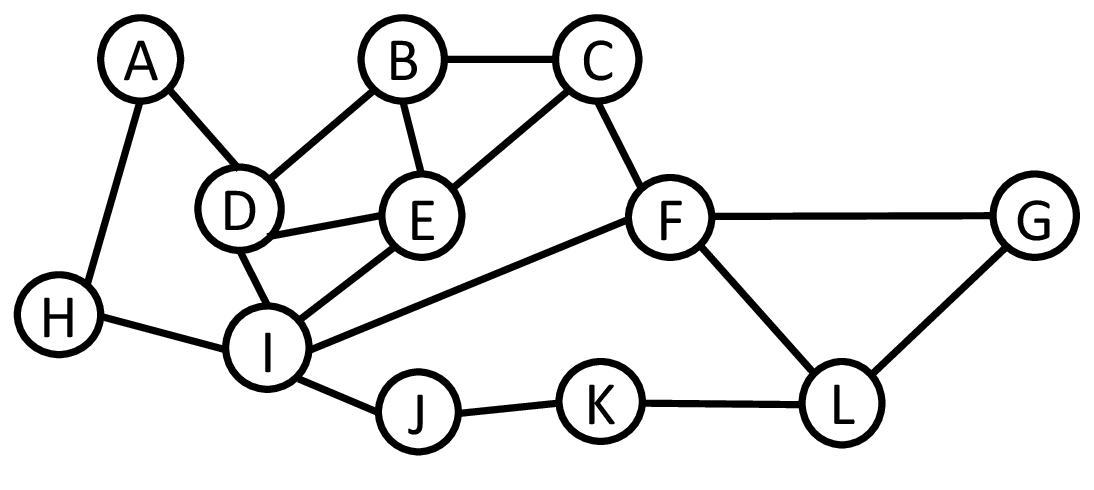}
    \caption{Original Graph $G$}
    \label{figure:originalgraph}
\end{subfigure}\hspace{0.008\textwidth}
\begin{subfigure}{0.32\textwidth}
    \includegraphics[width=\linewidth]{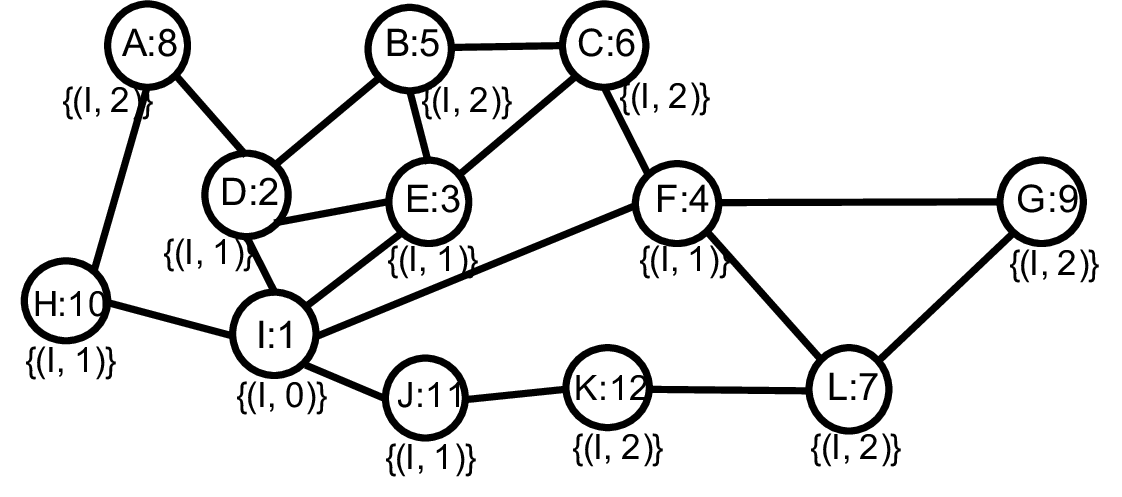}
\caption{Label I in $G$ with Vertex Rank}
  \label{figure:spreadI}
\end{subfigure}\hspace{0.008\textwidth}
\begin{subfigure}{0.32\textwidth}
    \includegraphics[width=\linewidth]{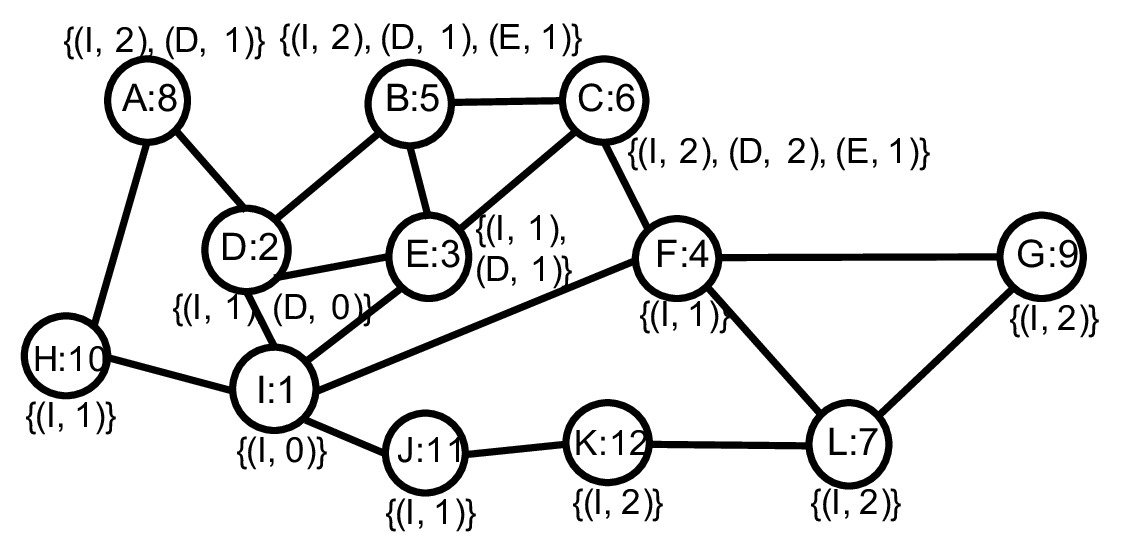}
    \caption{Label I, D, and E in $G$}
  \label{figure:spreadE&D}
    \end{subfigure}
    \caption{2-Hop Labeling and PLL Example}
    \label{figure:VCPLL-example}
\end{figure*}

\subsection{2-Hop Labeling and PLL} 
\label{preliminary}

The $2$-hop labeling algorithm~\cite{cohen2hop}, which was pioneered by Cohen {\em et al.}~\cite{cohen2hop}, provides an efficient scheme to answer distance queries. It assigns each vertex $u$ in the (undirected) graph a label $L(u)$ such that for any two vertices $u$ and $v$, their distance can be computed using only their label information. 
Formally, we compute $L(u)$ and for each  $h \in L(u)$,  the corresponding distance  from  $u$, i.e,  $d(h,u)$. 
It is also called {\em hub labeling} ~\cite{Abraham:2012:HHL:2404160.2404164} as the label set 
$L(u)$ for the vertex $u$ is referred to as the  hubs of $u$. 
Table~\ref{table:2hoplabel}  illustrates a $2$-hop labeling of the undirected graph $G$ (Figure~\ref{figure:originalgraph}).

Formally, the shortest path distance query $Dis(\cdot,\cdot)$ between any two vertices $u$ and $v$ can be answered as: 
$$ Dis(u,v) = \min_{h \in L(u) \cap L(v)} d(u,h)+d(h,v)$$
Thus, $2$-hop labeling can answer distance queries efficiently by traversing two lists of vertices, 
with an operation similar to merge sort.  

Given this, $2$-hop labeling aims to minimize the total labeling size, i.e,  if  $V$ is the vertex set of graph, the goal is to minimimize   $\sum_{u \in V} |L(u)|.$ 
For a directed graph, each vertex $v$ is labeled with two labels $L_{out}(u)$ (the hubs reachable from $u$) and $L_{in}(u)$ (the hubs reaching $u$) together with their distances to and from the  vertex $u$, and the objective function is minimizing  $\sum_{u \in V} |L_{out}(u)| +|L_{in}(u)|$.

The traditional approach employs an approximate (greedy) algorithm based on set-covering, which can produce a distance oracle with a size no larger than the optimal one by a logarithmic factor. 
Conceptually, the ground set consists of all the reachable vertex pairs, such as $\{(u,v): u \mbox{ reaches } v \}$.  
Any subset $C_v \subseteq S \times S$ ($S \subseteq V$) consisting of vertex pairs $(x,y)$ ($x,y \in S$), which can use $v$ to recover their shortest path distance, i.e., $d(x,y)=d(x,v)+d(v,y)$, is a candidate set. 
In other words, it suggests the benefits (effects) by assigning $v$ to the labels of vertices in $S$. 
The algorithm iteratively selects a vertex $v$ to label a subset of vertices $S$ and cover all the vertex pairs in $C_v \subseteq S \times S$. It continues until the entire ground set is covered. The criterion of selecting optimal $v$ and $S$ is based on the {\em ratio} of newly covered pairs in $C_v$, i.e., 
$|C_v \setminus P|$ where $P$ consists of already covered pairs, and the label cost $|S|$ : $\frac{|C_v \setminus P|}{|S|}$.  For any given vertex $v$, finding the optimal subset of vertices $S$ to be labeled is equivalent to a {\em densest subgraph} problem~\ref{table:2hoplabel}.     

The major problem with the set-cover based $2$-hop labeling approach is its high construction cost: its original complexity is as high as $O(n^5)$~\cite{cohen2hop}, which has then 
be reduced to  $O(n^3 \log n)$ with the latest optimization techniques~\cite{Babenko:2016:AHL:2997037.2996593}.  A number of other improvements under the set-cover framework   ~\cite{hopiedbt,ChengYLWY06,DBLP:conf/sigmod/JinXRF09,DBLP:conf/sigmod/JinRXL12,Thorup:2004:COR,DBLP:conf/edbt/ChengYLWY08,Abraham:2011:HLA,Geisberger:2008,Sankaranarayanan:2009} still cannot scale to the real world graphs that 
have millions or even  billions of vertices and edges.

\noindent{\bf Hierarchical Hub Labeling (HHL) and Canonical Hierarchical Hub Labeling (CHHL):}
An important direction to make $2$-hop labeling feasible and scalable for large graph is to restrict the choices of labeling (by imposing some special properties on what can be added into the labels).

\bdefin (Hierarchical Hub Labeling)
Given two distinct vertices $u$ and $v$, we say $u \succeq v$ if $u\in L(v)$ ($u$ is a hub of $v$). 
A hub (2-hop) labeling is {\em hierarchical} if  $\succeq$ forms a partial order. 
\edefin

In fact, any partial order can be extended to a total order (the order-extension principle) and for a set of vertices $V$, the total order is defined as a bijection $\pi: V \rightarrow {1, \cdots, |V|}$ ($\pi(v)$ is the rank of $v$). Given this, we can say that a label is hierarchical if there is a total order $\pi$ which satisfies: $u \in L(v)$ then $\pi(u) < \pi(v)$ ($u$ ranks higher than $v$). 

\bdefin (Canonical Hierarchical Hub Labeling) 
\label{def:CHHL}
Let the shortest path vertex set $P_{uv}$ consist of all vertices on shortest paths between $u$ and $v$ (including $u$ and $v$).
Given a total order $\pi$ on $V$, its canonical hub labeling is defined as follows: $u \in L(v)$ if $u$ has the highest order in $P_{uv}$, i.e., no other vertex $w$ in $P_{uv}$ such that $\pi(w)<\pi(u)$. 
\edefin

An important implication of canonical hierarchical hub labeling is that {\em it produces the minimal hierarchical hub labeling} for a given order~\cite{BGKSW2015}. Thus, the optimal HHL problem can be transformed into two sub-problems: 1) finding the optimal order that minimizes the label size; 2) computing the canonical HHL with respect to a given vertex order. 

A main breakthrough enabling efficient 2-hop labeling is the discovery of a simple, yet elegant algorithm called pruned landmark labeling (PLL) ~\cite{akiba2013fast}. It computes the canonical HHL (the second subproblem) for a given vertex order efficiently. Independently, essentially the same style algorithm was discovered for 2-hop reachability labeling, and is called {\em distribution labeling}~\cite{jin2013simple}. In the past few years, a number of studies~\cite{delling2014robust,li2017experimental} have further validated and confirmed the efficiency and effectiveness of PLL style algorithms for distance labeling.

Theoretically, the optimal hierarchical hub labeling (HHL) as well as the original 2-hop labeling have recently been proved to be NP-hard ~\cite{BGKSW2015}, which implies that the optimal order sub-problem (the first sub-problem listed above) is NP-hard as well.  A few heuristics, such as the ranking by degree and betweenness, have been developed for addressing this sub-problem~\cite{li2017experimental}. 
The second sub-problem (labeling generation)  typically dominates the overall labeling computation and is thus the focus of this study.




\begin{table}[t]
\scriptsize
      \centering
        \begin{tabular}{|c|l|}
            \hline
        Vertex&Labels\\ \hline
        \textit{A} &\{(\textit{A}, 0), (\textit{D}, 1), (\textit{I}, 2)\}    \\
        \textit{B} &\{(\textit{B}, 0), (\textit{D}, 1), (\textit{E}, 1), (\textit{F}, 2), (\textit{I}, 2)\}     \\
        \textit{C} &\{(\textit{C}, 0), (\textit{D}, 1), (\textit{E}, 1), (\textit{F}, 1), (\textit{I}, 2), (\textit{D}, 2)\} \\
        \textit{D} &\{(\textit{D}, 0), (\textit{I}, 1)\}     \\
        \textit{E} &\{(\textit{E}, 0), (\textit{D}, 1), (\textit{I}, 1)\} \\
        \textit{F} &\{(\textit{F}, 0), (\textit{I}, 1)\}    \\
        \textit{G} &\{(\textit{G}, 0), (\textit{F}, 1), (\textit{L}, 1)(\textit{I}, 2)\}    \\
        \textit{H} &\{(\textit{H}, 0), (\textit{A}, 1), (\textit{I}, 1)\}    \\
        \textit{I} &\{(\textit{I}, 0)\}     \\
        \textit{J} &\{(\textit{J}, 0), (\textit{I}, 1)\}     \\
        \textit{K} &\{(\textit{K}, 0), (\textit{J}, 1), (\textit{F}, 2), (\textit{I}, 2)\}     \\
        \textit{L} &\{(\textit{L}, 0), (\textit{F}, 1), (\textit{I}, 2)\}     \\
  \hline
    \end{tabular}
    \caption{2-hop labeling for Graph $G$ }
    \label{table:2hoplabel} 
\end{table}

\subsubsection{Pruned Landmark Labeling (PLL)}
Given a total order $\pi$ of vertices, the pruned landmark labeling  algorithm (PLL) ~\cite{akiba2013fast} assigns each vertex, based on the order $(\pi(v_1) <\pi(v_2)< \cdots <\pi(v_n))$, to the labels of other vertices in the graph following a BFS process.
As it assigns the vertex $u$ with rank $\pi(u)$ to a vertex $v$ with lower rank $(\pi(u)<\pi(v))$, it needs to check if $u$ is the highest rank vertex in the shortest paths between $u$ and $v$ ($P_{uv}$). This is the canonical HHL condition and can be done by determining {\em whether the distance between $u$ and $v$ can be recovered by a certain higher ranking vertex}: $$d(u,v) < d(v,h)+ d(h,u), \mbox{ for all } h \in L(u) \cap L(v).$$ 
When the condition does not hold, $u$ will be pruned by $v$ (i.e., is not added into the label of $v$ and will not further expand from $v$) during the labeling process. 

\begin{algorithm}[t]
\caption{PLL for $G=(V,E)$ with Order $\pi$}
\label{alg:DLD}
\begin{small}
\begin{algorithmic}[1]
\FORALL {$u \in V$ \COMMENT{following order $\pi$ from high to low}}
    \STATE Queue $Q=\{(u,0)\}$ \COMMENT{BFS process to use $u$ for labeling}
    \WHILE{$Q$ is not empty}
        \STATE $(v,d(u,v)) \leftarrow$ $Q$.pop()
        \IF{$d(u,v) < \min_{h \in L(u) \cap L(v)} d(u,h)+d(h,v)$}
          \STATE Add $(u, d(u,v))$ into $L(v)$ 
          \STATE For all $v^\prime$ of $v$'s neighbor when $v^\prime$ unvisited by $u$ and $\pi(u)<\pi(v^\prime)$,  Add $(v^\prime,d(u,v)+1)$ to $Q$
         \ENDIF
      \ENDWHILE
 \ENDFOR
\end{algorithmic}
\end{small}
\end{algorithm}

Algorithm~\ref{alg:DLD} sketches the labeling process for an undirected graph. Note that $d(u,v)$ in the algorithm is the distance computed by the BFS process, which may not be the exact distance between $u$ and $v$ (due to the pruning effect). But the  recorded distance  in the label (Line $6$)  is always exact (since it can travel through all the shortest paths starting from $u$ reaching to $v$).

\noindent{\bf Small Revision:} In Algorithm~\ref{alg:DLD}, which is  slightly different from  the standard BFS process as well as all the previous  PLL descriptions~\cite{akiba2013fast,li2017experimental} the 
following change is made. In Line $7$, {\em we only send the current labeling vertex $u$ to the neighbors of $v$ that have  rank lower than $v$ ($\pi(u)<\pi(v^\prime)$)}~\cite{BGKSW2015}. This can reduce the cost of sending 
$u$ to $v^\prime$ if $v^\prime$ has higher rank than $u$. Based on the canonical labeling criterion (Definition~\ref{def:CHHL}), $u$ cannot be added to $v^\prime$ and can be safely pruned without further expansion from $v^\prime$. 

\noindent{\bf Running Example:} Figures ~\ref{figure:spreadI} and ~\ref{figure:spreadE&D} illustrate the first three vertices $I, E$, and $D$ of the PLL labeling process for graph $G$ (Figure~\ref{figure:originalgraph}) with its order explicitly denoted in  Figure~\ref{figure:spreadI} and Figure~\ref{figure:spreadE&D}. For instance, $I$ is ranked first and $D$ is ranked second, and so on.

\subsection{Vertex Centric Graph Computing Models}
\label{subsection:VC}

The seminal vertex-centric programming model proposed by the Pregel paper~\cite{Malewicz:2010:PSL} is one of the key driving forces behind recent parallel graph processing system research~\cite{gonzalez2012powergraph,kyrola2012graphchi,nguyen2013lightweight,shun2013ligra,low2014graphlab,khorasani2014cusha,zhang2015numa,sengupta2015graphreduce,wang2016gunrock,pai2016compiler,maass2017mosaic,han2017graphie,dathathri2018gluon}. It is also known as the ``think-like-a-vertex" model. Though other models have been used,  the  simplicity, wide-range applicability, and strong scalability make the vertex-centric model  very appealing as the basic interface and abstraction for parallel graph processing~\cite{mccune2015thinking}. 

Simply speaking,  parallel graph processing is viewed as an iterative process, where each iteration traverses/processes the {\em active} vertices -- on 
each vertex, we perform  computations based on the data from  incoming and/or outgoing edges together with the local vertex data, and then update the values/state  associated with  the vertex.   
The vertices that record a change in their local state become the  active vertices for the next iteration. 
The parallelization typically uses the Bulk Synchronous Parallel (BSP) execution~\cite{valiant1990bridging} and requires  a global synchronization at the end of each iteration.
The entire process terminates once the set of active vertices  becomes empty.

\begin{algorithm}[t]
\caption{Vertex-Centric (Scatter-Gather) Computation (G=(V,E))}
\label{alg:VC-SG}
\begin{small}
\begin{algorithmic}[1]
\STATE Initialize ActiveVertices $\subseteq$ V
\WHILE{ActiveVertices is not empty}
\STATEx \COMMENT{Scatter Phase:}  
\FORALL{$a \in$ ActiveVertices} 
    \STATE a.Scatter(a.edges) : \COMMENT{for each edge $e=(a,v)$ of $a$, send message($a,e,v$) to $v$} 
\ENDFOR
\STATEx ActiveVertices $\leftarrow \emptyset$
\STATEx \COMMENT{Gather Phase:}
\FORALL{$v$ Received Message}
    \STATE v.Gather($v$.messages): \COMMENT{vertex compute using received messages and update its value} 
    \STATE ActiveVertices $\leftarrow$ $\{v: \mbox{v.value is updated} \}$
\ENDFOR
\ENDWHILE
\end{algorithmic}
\end{small}
\end{algorithm}

In this paper, we will focus on studying how PLL can be parallelized under the vertex-centric computation. 
A high level abstraction of the vertex-centric computation based on a scatter-gather model~\cite{Malewicz:2010:PSL,roy2013x} is sketched in Algorithm~\ref{alg:VC-SG}. 
Each vertex computation is described through  two functions: 1) the  {\em Scatter} function, which describes how each vertex uses its vertex value and edge value to propagate a message to its neighbors; and 2) {\em Gather} function, which describes how each vertex computes a new value based on its original value and all the new messages it received. 
Each phase can traverse in parallel their corresponding vertex sets: {\em ActiveVertices}, including the vertices need to send out messages to their neighbors, for scatter phase, and vertices that  have  received a new message to be processed in the scatter phase. When a vertex is updated with a new value (in the {\em Gather} function), it will be added to the set ({\em ActiveVertices}). The process continues until there are no new active vertices. 

Various more advanced parallel graph programming models are proposed to further refine the vertex-centric model. This 
includes the  GAS (Gather-Apply-Scatter)~\cite{gonzalez2012powergraph} and the push and pull models~\cite{patterson2012direction,shun2013ligra,besta2017push}, where the goal is to better fit the computational and communication patterns of graph processing. 
There is also work on generalizing the model to finer granularity, such as the edge-centric model~\cite{roy2013x}, or to coarser granularity, such as path or subgraph-centric~\cite{quamar2016nscale}, and $k$-step neighborhood~\cite{cao2015grarep,ko2018turbograph++} models. However, these models do not necessarily provide more advantage/capability to  support the parallelization of the PLL than the aforementioned vertex centric model. 
Other recent efforts like iBFS~\cite{liu2016ibfs}, CUBE~\cite{zhang2016exploring}, and RStream~\cite{wang2018rstream} target different applications or algorithms, and have distinct challenges associated with them.

\section{Basic Vertex-Centric  Algorithm}\label{design-vc}




Recall that the PLL algorithm (Alg.~\ref{alg:DLD})  iterates following the vertex rank (order): at the 
$i^{th}$ iteration, the vertex $u$ with rank $\pi(u)=i$ will be distributed to all other vertices in the graph using a BFS process. The key condition to add $u$ into the label of $v$, $L(v)$, is the distance check for the {\em canonical} labeling criterion: the distance between $u$ and $v$ cannot be recovered by earlier processed vertices, i.e., vertices with rank higher than $u$: $d(u,v) < d(u,h)+ d(h,v)$ for any $h \in L(v) \cap L(u)$ and $\pi(h)<\pi(u)$. Otherwise, $u$ will not be assigned to $v$ and will not be sent to $v$'s neighbors for any further expansion. 

The main challenge in  parallelizing PLL is that adding a vertex $u$ of rank $\pi(u)$ to another vertex $v$ in the BFS traversal seems to be {\em dependent} on the completion of labeling of all higher ranked vertices (i.e., any vertex $h$ such that $\pi(h)< \pi(u)$) in order to apply  the distance check. In comparison,  for parallelization with  the vertex-centric model, we would like to distribute all vertices to their neighbors simultaneously for vertex labeling. 
Given this, {\em the need to distribute all vertices simultaneously and the distance checking condition based on the vertex rank} seems to be in conflict as there is no guarantee that  the higher vertices can finish the  distribution before lower rank ones.  Indeed, as we mentioned earlier, all the existing attempts have all failed to parallelize inter-vertex labeling while preserving the canonical labeling criterion~\cite{akiba2013fast,ferizovic2015parallel,qiu2018parapll}.


\subsection{The Algorithm}
The main insight to help us solve the aforementioned dilemma is as follows. Assume we spread all vertices simultaneously into the graph (starting by sending each vertex to their neighbors), and we do the spreading iteration by iteration following the vertex-centric programming model. Let us consider a  vertex $u$ with  the  rank $\pi(u)$ that reaches the  vertex $v$ at the $j$-th iteration: clearly in order to determine if $u$ should be added to  $L(v)$, and spread by $v$ continuously, we need to verify if there are any other vertices, say $w$,  with higher ranks than $u$ ($\pi(w)<\pi(u)$), which can produce an equal or shorter distance, i.e., $d(u,v) \geq d(u,w)+d(w,v)$. Note, our key insight is that  {\em if such vertex $w$ exists for the testing, then, it must be able to reach both $u$ and $v$ within the $j$-th iteration ($d(u,v)$ steps)}.

In retrospect, the distance check condition for canonical labeling criterion requires not only the labeling of higher ranked vertices $h$ to be completed before the distance check between $u$ and $v$, but also their distances $d(u,h)$ and $d(h,v)$ to be smaller than $d(u,v)$. The latter condition is the key for utilizing a VC model for PLL, and provides {\em a natural match of the vertex spreading process at the heart of VC computation to the center mechanism of the canonical labeling in PLL}: 
If we follow a basic label spreading process in VC, then, we can in parallel prune (or accept) vertex labels at any vertex using the distance check for canonical labeling in PLL.


\begin{algorithm}[t]
\scriptsize
\caption{VC-PLL for $G=(V,E)$ with Order $\pi$}
\label{alg:VertexCentricPLL_push}
\begin{algorithmic}[1]
\STATEx \COMMENT{Init.: ($L(v)$: label; $\delta L(v)$: new label from each iteration)}
\STATE $ActiveVertices \leftarrow V; \forall v \in V, \delta L(v) \leftarrow \{(v, 0)\}, L(v) \leftarrow \delta L(v)$ 
\WHILE{$ActiveVertices \ne \emptyset$}
\STATEx \COMMENT{Scatter Phase:}  
\FORALL{$a \in$ ActiveVertices} 
    \STATEx \ \ \ \ \ a.Scatter(a.edges): 
    \FORALL{$(a,v) \in a.edges$}
         \STATE for all $(u,d(u,a)) \in \delta L(a)$, when $\pi(u) < \pi(v) \wedge u \notin L(v)$: send $(u, d(u,a)+1)$ to $v.messages$  \label{alg:VertexCentricPLL_push:line:7}
    \ENDFOR 
\ENDFOR
\STATEx ActiveVertices $\leftarrow \emptyset$
\STATEx \COMMENT{Gather Phase:}
\FORALL{$v \in V: v.messages \neq \emptyset$\COMMENT{Received Message}}
    \STATEx \ \ \ \ \ v.Gather($v$.messages):  
    \STATE $\delta L(v) \leftarrow \emptyset$
	\FORALL {unique $(u, d(u,v)) \in v.messages$}
		 \label{alg:VertexCentricPLL_push:line:15}
		\IF {$d(u,v) < \min_{h \in L(u) \cap L(v)} d(u,h)+d(h,v)$} 
		    \STATE Add $(u,d(u,v))$ to  $\delta L(v)$
        \ENDIF
	\ENDFOR
	\STATE $\delta L(v) \neq \emptyset$: $L(v) \leftarrow L(v) \cup \delta L(v)$; Add $v$ to ActiveVertices
\ENDFOR
\ENDWHILE
\end{algorithmic}
\end{algorithm}

\noindent{\bf Algorithm Description:}
Algorithm~\ref{alg:VertexCentricPLL_push} sketches the main process of performing PLL based on the vertex-centric computation model (Algorithm~\ref{alg:VC-SG}). In the {\em Initialization} phase, all vertices are active initially ($ActiveVertices=V$). For each vertex $v$, $L(v)$ records the partial label and $\delta L(v)$ records the new label being generated at each iteration. Initially, both labels of $v$ records itself and distance $0$ (any vertex reaches itself in zero steps). The main computation alternates between the {\em Scatter} phase and {\em Gather} phase and will continue until no new active vertices exist (Lines $2$ to $17$): 

\noindent{\bf 1) {\em Scatter} phase} (Lines   $3$ to $7$, also referred to as the push model):  all active vertices with new labels perform a vertex {\em Scatter} function (Lines $4$ to $6$): 
each sends their new vertex labels with the updated distance: $(u,d(u,v)) \in \delta L(u)  \rightarrow (u,d(u,v)+1)$ to all their neighbors (Line $5$) with two conditions: the rank of vertex $u$ needs to higher than $v$ (otherwise, it will be pruned) and it has never been added to the label of $v$.  


\noindent{\bf 2)  {\em Gather} phase} (Line $8$-$16$): all vertices that receive a new message ($v.messages \neq \emptyset$) perform a vertex {\em Gather} function (Lines $9$-$15$): For a vertex $v$, it traverses all its received messages (distance label from its neighbors), and for each {\em unique} vertex $(u,d(u,v))$ across the set of 
messages, 
it confirms the distance check for the canonical labeling criterion: for a distance label message $(u,d(u,v))$, $d(u,v)$ must be smaller than the distances via any existing labels ($L$), i.e., $d(u,v)< \min_{h \in L(u) \cap L(v)} d(u,h)+d(h,v)$ (Line 11). If this true, it will be added into $\delta L(v)$. Once $\delta L(v)$ is computed and it is not empty, we will add it into $L(v)$ and add $v$ to ActiveVertices (Line $15$).  
Note that we need to identify unique vertices in the step above, because two neighbors may send the same vertex $u$. 




\begin{figure*}[th]
\begin{subfigure}{0.32\textwidth}
    \includegraphics[width=\linewidth]{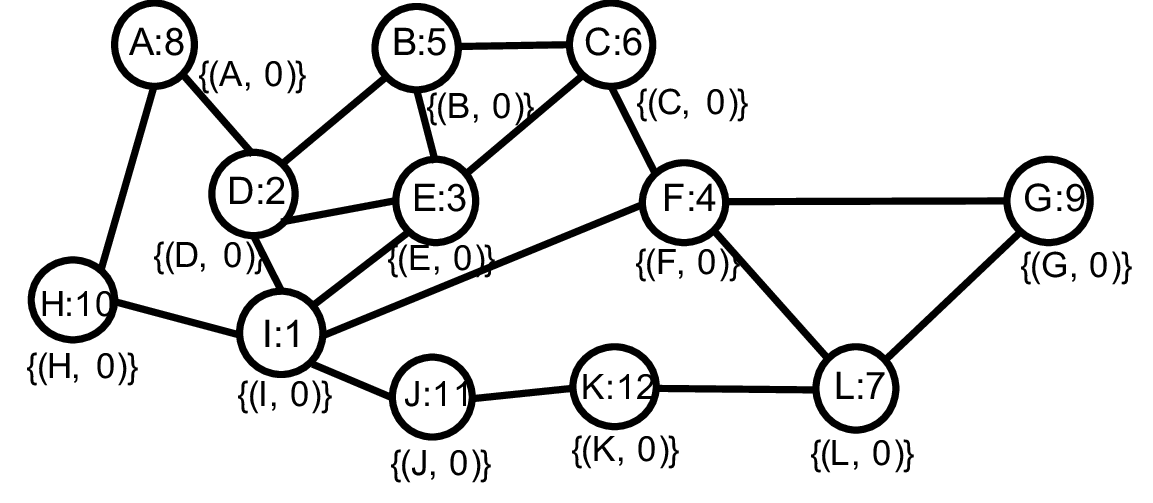}
    \caption{Initial \textbf{$\delta L$} of Graph $G$}
  \label{figure:VCPLL0}
\end{subfigure}\hspace{0.008\textwidth}
\begin{subfigure}{0.32\textwidth}
    \includegraphics[width=\linewidth]{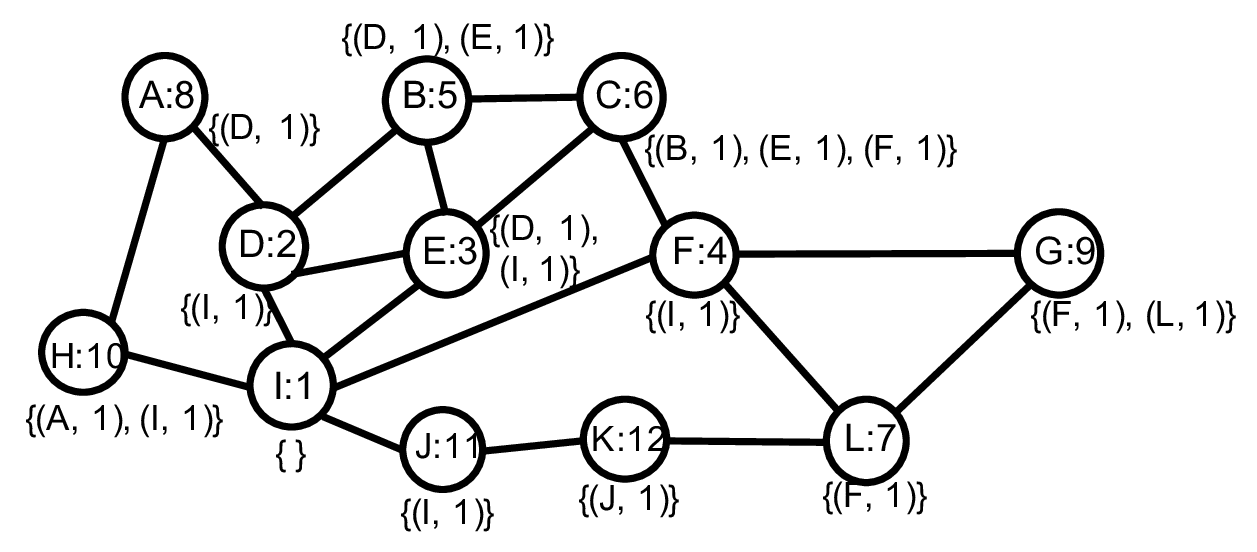}
    \caption{$\delta L$  for Graph $G$ after Iteration 1}
  \label{figure:VCPLL1}
\end{subfigure}\hspace{0.008\textwidth}
\begin{subfigure}{0.32\textwidth}
    \includegraphics[width=\linewidth]{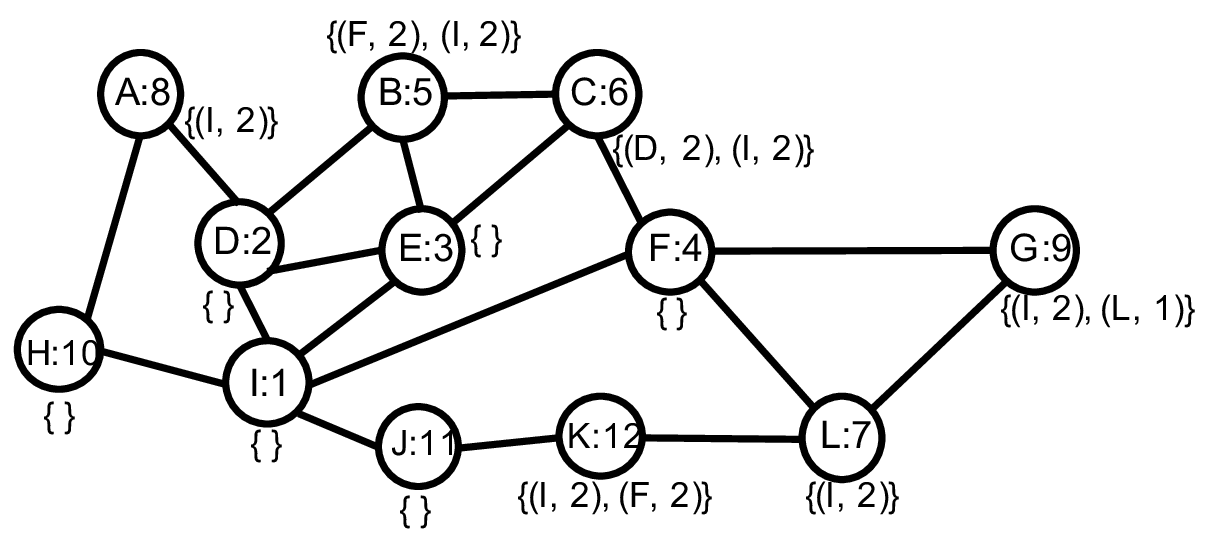}
    \caption{$\delta L$ for Graph $G$ after Iteration 2}
  \label{figure:VCPLL2}
    \end{subfigure}
    \caption{A VC-PLL Running Example}
    \label{figure:VCPLL-example}
\end{figure*}

\noindent{\bf Running Example:} Figures~\ref{figure:VCPLL0}  illustrates the iteration of label spreading, where the labels in the graph record newly generated labels $\delta L$ for all vertices. At each iteration, $L(v)$ is simply the union of all $\delta L(v)$ from all earlier iterations.

\subsection{Theoretical Properties}
\noindent{\bf Correctness:}
Theorem~\ref{thm:correct} proves that VC-PLL produces a canonical hierarchical labeling and therefore also generates the minimum labeling size given a vertex order. In other words, VC-PLL produces the same label as the original PLL.

\bthm
\label{thm:correct}
VC-PLL (Algorithm~\ref{alg:VertexCentricPLL_push}) produces the canonical hierarchical hub labeling given a vertex order $\pi$. 
\ethm

\bproof 
Recall the shortest path vertex set $P_{uv}$ consists of all vertices on shortest paths between $u$ and $v$ (including $u$ and $v$). Then, we need to prove {\em $u \in L(v)$ iff $u$ has the highest order in $P_{uv}$ (Definition~\ref{def:CHHL})}.

\begin{wrapfigure}{r}{0.5\textwidth}
  \begin{center}
    \includegraphics[width=0.15\linewidth]{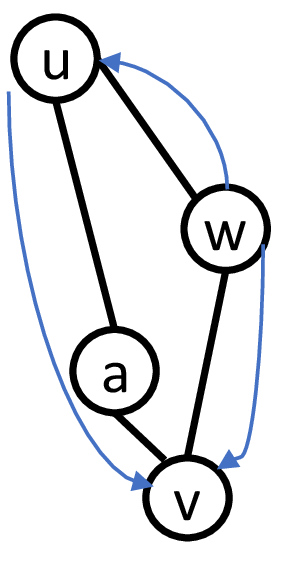}
  \end{center}
  \caption{}
  \label{figure:correct}
\end{wrapfigure}

First ($\rightarrow$), we can see that if $u \in L(v)$, then we cannot find another vertex $w$ with rank higher than $u$, such that $d(u,v) \geq d(u,w)+d(w,v)$. Thus, $u$ must have highest order in $P_{uv}$.
If not, assume we have another vertex $w \neq u$ that  has the highest rank in $P_{uv}$. Then, based on our algorithm, $w$ will be the highest ranked in $P_{wu}$ and $P_{wv}$. Thus, $w$ can always reach $u$ and $v$ before $u$ reaches $v$ (Figure~\ref{figure:correct}) and it is in $L_{u}$ and $L_{v}$ when $u$ reaches $v$.

Second ($\leftarrow$), assuming $u$ has the highest order $P_{uv}$, then, based on the same argument, it can definitely go through a shortest path from $u$ to $v$ using Algorithm~\ref{alg:VertexCentricPLL_push} and if it reaches $v$, no other vertices in $L_v$ (and $L_u$) can prune it.
\eproof

The following corollary can be immediately obtained. 
\bcoroll
\label{col:labelcoroll}
In VC-PLL, when a distance label $(u,d(u,v))$ is added into $\delta L(v)$, $d(u,v)$ is the exact shortest path distance between $u$ and $v$, and $u$ has the highest rank in $P_{uv}$. Further, at any time $L(v)\subseteq {\mathcal L}(v)$, where ${\mathcal L}(v)$ is the final complete label of $v$.
\ecoroll

\noindent{\bf Tree Width and Time Complexity} 
Following the approach in PLL~\cite{akiba2013fast}, we can obtain a theoretical upper-bound of VC-PLL's time complexity. 

\bthm 
\label{thm:treewidthtimecomplexity}
Assuming graph $G$ with a tree-decomposition~\cite{ROBERTSON1986309} of tree-width $w$, then there is a vertex order $\pi$, in which the VC-PLL takes $O(w|E| \log |V|+ w^2 |V|(\log |V|)^2)$ time (the same as that of PLL~\cite{akiba2013fast}). 
\ethm
Proof is omitted due to space limitation. 



\subsection{Limitations and Benefits of VC-PLL}
\label{subsection:limitations}
We note that even though Theorem~\ref{thm:treewidthtimecomplexity} provides a theoretical evidence on time complexity, it does not provide a direct comparison of the computational and memory access costs between these two algorithms, PLL and VC-PLL, for a given vertex order. In the following, we will do an in-depth comparison between VC-PLL and PLL, and identify the main performance bottleneck and potential benefits introduced in VC-PLL.  
Since the cost of generating (sending) distance labels and distance check dominates the total computation (similar in the original  PLL~\cite{akiba2013fast}), we will primarily focus on these two factors for computational costs. In addition, we will compare the memory access of the underlying graph $G$ between them. 

\noindent{\bf Additional Cost of Distance Label Generation:}
For a given vertex $u$, PLL will send it to a vertex $v$ only once. In BFS, PLL will flag $v$ after one distance label $(u,d(u,v))$ is passed through (Line $7$ in Algorithm~\ref{alg:DLD} is sequentially executed). But VC-PLL can send multiple $(u,d(u,v))$ messages to the same $v$ at two consecutive iterations. 

\blemma
\label{lemma:messagefiltring}
Given vertex $u$ and vertex $v$, a distance label $(u:d(u,v))$ may reach $v$ at exactly two possible and consecutive iterations: 
Let $a$ be a neighbor of $v$, and $u \in L(a)$ ($u$ is the highest rank vertex in $P_{ua}$), then it reaches $v$ at $d(u,a)+1$ iteration, which is either: 1) equal to the shortest path distance between $u$ and $v$, and $u$ may or may not be added to $L(v)$; or 2) equal to $d(u,v)+1$, i.e., the path from $u$ to $a$ to $v$ is one step longer than the shortest path between $u$ and $v$, and $u$ will be pruned. 
\elemma 
Please refer to Appendix for proof. 

In addition, at each of these two iterations, if $u$ has not been or is not in the label of $v$, then different neighbors of $v$ may send the same $(u,d(u,v))$ messages to $v$. 

\noindent{\bf Additional Cost of Distance Check:}
\blemma 
\label{lemma:labelmessageseset}
The set consisting of all pairs $(u,v)$ for distance checks is the same in PLL and VC-PLL. 
\elemma 
Please see Appendix for proof. 

However, {\em the number of distance checks in VC-PLL can be higher than PLL, as a vertex $u$ can be sent to $v$ in two consecutive iterations in VC-PLL}. 

The computational cost of distance check 
$$d(u,v) < \min_{h \in L(u) \cap L(v)} d(u,h)+d(h,v)
$$ in VC-PLL is also higher than that in $PLL$. In VC-PLL, the cost is $O(|L(u)|+|L(v)|)$, where $L(u)$ and $L(v)$ are (partial) labels of $u$ and $v$ at the time of distance check for $d(u,v)$. Assuming $L(u)$ and $L(v)$ are not sorted, we can first map $L(u)$ into an array or hash-table, and then check all the vertices in $L(v)$ against the above data structure. 
In PLL~\cite{akiba2013fast}, since we process vertex $u$ one at a time, and when we try to process $u$, its label $L(u)$ is already computed. Thus, we can first map $L(u)$ to an array only once at the beginning of the BFS iteration. Thus, the cost of $O(|L(u)|)$ can be practically saved for each distance check; thus the distance check for PLL is only $O(|L(v)|)$. For VC-PLL, we cannot do this directly as it is prohibitively expensive to map every $L(u)$ to an array or hash-table at the same time.

To summarize,  VC-PLL introduces redundant distance labeling messages, which may also lead to redundant distance checks $d(u,v)$. Furthermore, individual distance checks in PLL can be much faster due to the reuse of $L(u)$ in an array or hash-table representation. In fact, these performance issues seems to challenge the capabilities of Vertex Centric (VC) computational in supporting: 1) effective message filtering and communication and 2) efficient remote (global) memory access. 

\noindent{\bf Reduced Memory Access Cost for Graph Topology:}
A potential benefit of VC-PLL is that it can help reduce the total memory access for the graph topology compared with PLL. Specifically, this is the total number of edge access in the graph (Line $7$ in PLL and Line $4$ in VC-PLL) for propagating distance messages: 1) For PLL, for each new vertex label message, an edge access $(v,v^\prime)$ is performed for adding $(v^\prime, d(u,v^\prime)+1)$ to the queue $Q$ --  thus, the number of total edge accesses is equivalent to  the number of total distance labeling messages. 2) For VC-PLL, for vertex $a$, all its new potential labels at an iteration $\delta L(a)$ are  filtered and grouped together for one edge access $(a,v)$.  Thus, VC-PLL should have less total memory access cost for the edges in graph than PLL (assuming they propagate similar number of labeling messages).

Also, in terms of an upper-bound estimation, following the assumption in Theorem~\ref{thm:treewidthtimecomplexity}, there is a vertex order which has $O(w \log |V| |E|)$ complexity for the edge cost in PLL ($w$ is the tree-width of $G$), whereas VC-PLL is bounded by $O(D |E|)$, where $D$ is the diameter of the graph. 
In the real-world graphs, the diameter of a graph is typically much smaller than its tree-width $w$  ~\cite{doi:10.1080/15427951.2016.1182952}.
Finally, we note such memory access benefits to be similar to ``frontier sharing'' in iBFS~\cite{liu2016ibfs}, though the latter is not based on vertex-centric computation. 

\noindent{\bf Sequential Performance Comparison:} We implemented Alg.~\ref{alg:VertexCentricPLL_push} (VC-PLL) and tested its performance on the {\tt DBLP} graph (Section~\ref{sec:eval}) against PLL using a single thread. We found that it has poor performance with a total execution time of $13,583$ seconds compared to less than $100$ seconds for  PLL!
It does not fare well against PLL in other graphs either. 
Basic performance analysis shows that the additional computational costs significantly outweigh the benefits of memory access cost reduction.

Now, the question we face is: {\em can VC-PLL overcome its limitations  and reduce those additional costs (message passing and remote memory access)?} In the next Section, we will discuss how we can extend the basic VC model to help achieve this and show that the new VC-PLL can be even faster than PLL  sequentially -- as it has less computational  as well as memory access costs.

\section{Batched Vertex-Centric Alg.}\label{sec:design-bvc}

Though VC-PLL can be described in a natural Vertex-Centric computational scheme, it also demonstrates certain limitations of the original vertex-centric model assumptions:
1) Typically, the vertex value (and message) is fixed in VC, whereas in VC-PLL, each vertex value (and message) is a continuously growing list (or set); 
2) In the Gather function, the computation needs remote memory access for checking distance conditions (Line 11 in Algorithm ~\ref{alg:VertexCentricPLL_push}): in most cases, $u$ is not a neighbor of $v$, and when we use $L(u)$ for distance checks, the memory access is remote with respect to vertex $v$.
Indeed, the additional computational costs of VC-PLL compared with PLL (Subsection~\ref{subsection:limitations}) 
can be traced back to these limitations.




\subsection{Batched Vertex-Centric Computation}
To deal with the performance inefficiency of VC-PLL and the limitations of the vertex centric computation model, we introduce a batched strategy for the standard VC computation. 
Batches are processed in sequence with the vertices within each batch being processed using the vertex centric computation. 
The batched strategy naturally introduces mechanisms to help handle: 1) (continuously increasing) size of vertex value and redundant message passing, and 2) remote vertex memory access.


\noindent{\bf Using Bit Operation for Efficient Message Passing and Filtering:}
In each batch processing step, an active vertex only processes up to $batch\_size$ unique labels. Based on this important observation, we can use a compact bit-vector data structure called {\em candidate bit-vector} for efficient message filtering. 
The basic idea is as follows.
Each active vertex maintains a candidate bit-vector with the length of $batch\_size$ bits, each bit corresponding to a vertex in the batch (e.g., if the $batch\_size$ is 1K, such candidate bit-vector is only 128 bytes). 
If a vertex $u$ in the current batch is sent to a vertex $v$, then its corresponding bit in the candidate bit-vector of $v$ is set. Note that the use of bit-vectors also allows  atomic {\em compare-and-swap} operation in the shared memory setting. Note that without batch processing, we have to consider doing an expensive list merge for handling message passing and aggregation (as the scatter and gather functions in VC-PLL for distance label messaging and processing, respectively).

\noindent{\bf Improving Data Locality for Remote Vertex Memory Access:} 
Simply speaking, only the  vertices in the current processing batch can be accessed remotely during the vertex-centric computation. Because the number of vertices in each processing batch is limited, we can use a compact data structure such as an array or hash-table to store their labels for efficient $O(1)$ access (similar to what is done in PLL for each processed vertex in distance checks). 


\subsection{BVC-PLL Algorithm}

\begin{algorithm}[t]
\scriptsize
\caption{BVC-PLL for $G=(V,E)$ with Order $\pi$}
\label{alg:BatchedVertexCentricPLL_push}
\begin{algorithmic}[1]
\STATEx \COMMENT{Init.: ($L(v)$: label; $\delta L(v)$: new label from each iteration)}
\STATE $\forall v \in V, L(v) \leftarrow 
\emptyset, C(v) \leftarrow \emptyset$ 
\STATE Split $V$ into equal-size batches: $B_1$, $B_2$, $\cdots$ $B_T$ where $B_i$ include the vertices with rank $(i-1) \times |V|/T +1$ to $i \times |V|/T$
\FORALL {$B_i:\ i=1\ to\ T$ \COMMENT{Labeling in Batch}}
\STATE $ActiveVertices \leftarrow B_i; \forall u \in B_i, \delta L(u) \leftarrow \{(u, 0)\}, L(u) \leftarrow L(u) \cup \delta L(u)$, and map $L(u)$ to Hashtable $H(u)$
\WHILE{$ActiveVertices \ne \emptyset$}\label{line:vcstart}
\STATEx \COMMENT{Scatter Phase:}  
\FORALL{$a \in$ ActiveVertices} 
    \STATEx \ \ \ \ \ a.Scatter(a.edges): 
    \FORALL{$(a,v) \in a.edges$}
         \STATE for all $(u,d(u,a)) \in \delta L(a)$, when $\pi(u) < \pi(v) \wedge u \notin C(v)$: flag $u$ in $C(v)$ and send $(u, d(u,a)+1)$ to $v.messages$  \label{alg:VertexCentricPLL_push:line:7}
    \ENDFOR 
\ENDFOR
\STATEx ActiveVertices $\leftarrow \emptyset$
\STATEx \COMMENT{Gather Phase:}
\FORALL{$v \in V: v.messages \neq \emptyset$ \COMMENT{Received Messages}}
    \STATEx \ \ \ \ \ v.Gather($v.messages$):  
    \STATE $\delta L(v) \leftarrow \emptyset$
	\FORALL {$(u, d(u,v)) \in v.messages$}\label{line:lcstart}
		 \label{alg:VertexCentricPLL_push:line:15}
		\IF {$d(u,v) < \min_{h \in L(u) \cap L(v)} d(u,h)+d(h,v)$} 
		    \STATE Add $(u,d(u,v))$ to  $\delta L(v)$
        \ENDIF
	\ENDFOR\label{line:lcend}
	\STATE $\delta L(v) \neq \emptyset$: $L(v) \leftarrow L(v) \cup \delta L(v)$; Add $v$ to ActiveVertices
	\STATE If $v \in B_i$: Add $\delta L(v)$ to $H(v)$
\ENDFOR
\ENDWHILE\label{line:vcend}
\STATE $\forall v \in V, C(v) \leftarrow \emptyset$\label{line:clear}
\ENDFOR
\end{algorithmic}
\end{algorithm}



Algorithm~\ref{alg:BatchedVertexCentricPLL_push} sketches the batched Vertex-Centric algorithm for PLL, referred to as BVC-PLL. Specifically, here, the batches of the vertices are formed according to the rank of each vertex (Line $2$). The earlier processed batch consists of the vertices with higher ranks (Line $3$). BVC-PLL labels vertices one batch at a time and for assigning the labels in each batch, the vertex centric computation in VC-PLL is  followed (Lines 5-21) -- more specifically,  the Scatter Phase and Scatter function, Gather Phase and Gather function is  preserved with only minor revisions for dealing with message passing and remote memory access. 

Each vertex $v$ is associated with a candidate-bit vector $C(v)$. Its length is equal to the batch size. It will be initialized for each batch (Lines $1$ and $22$). 
During the Scatter phase, for any vertex $a$ to send a message $(u,d(u,a)+1)$ to its neighbor $v$, it will check if $u$ is sent to $v$ before ($u \notin C(v)$, Line $8$). This corresponds to the {\em unvisited} flag in the original PLL. Due to the atomic compare-and-swap operation, it can guarantee only one message from $u$ is being sent to $v$ and thus help resolve the redundant distance labeling generation problem (in Subsection~\ref{subsection:limitations}). 

Each vertex $u$ in the batch $B_i$ will map its existing label $L(u)$ to a hash-table (or array) $H(u)$ at the beginning of vertex-centric computation (Line $4$). Since the new label of $u$ may be generated during the labeling process, we will map the new label $\delta L(v)$ to $H(v)$ when the update is available (Line $19$). Given this, the distance check (in Line $14$) only needs to go through $L(v)$, and thus has the same distance check cost as the original PLL (Subsection~\ref{subsection:limitations}). 

\noindent{\bf Correctness:}
It is easy to see that {\em BVC-PLL (Algorithm~\ref{alg:BatchedVertexCentricPLL_push}) produces the canonical hierarchical hub labeling given a vertex order $\pi$}: the canonical labeling criterion ($u \in L(v)$  if $u$ has the highest rank in $P_{uv}$) is maintained as BVC-PLL can assign $u$ to $L(v)$ at $u$'s batch correctly  (Theorem~\ref{thm:correct}) following the batch processing order. 

Another interesting property is that when the batch size reduces to one, i.e., when we process one vertex at a time, then BVC-PLL behaves exactly the same as the original PLL~\cite{akiba2013fast}.

Finally, we note that introducing and using  bit-vector $C(v)$ for each vertex $v$ and $H(u)$ for each processing batch vertex $u$ does not introduce additional time complexity compared with PLL. PLL uses only one bit for each vertex $v$ as the {\em visited} flag and one $H(u)$ for distance check, whereas BVC-PLL simply utilizes a group of them at the same time. Thus, the time complexity results of Theorem~\ref{thm:treewidthtimecomplexity} hold for BVC-PLL as well.   

\subsection{Detailed Computational Cost Comparison}
\label{subsection:apple2apple}
In the following, we provide an apple-to-apple computational cost analysis between BVC-PLL and PLL. Following Subsection~\ref{subsection:limitations}, we will focus on the cost of generating (sending) distance labels and distance checks. 

\noindent{\bf Cost of Distance Label Generation:}
Since in BVC-PLL, each vertex $u$ can be sent to $v$ exactly once, together with Lemma~\ref{lemma:labelmessageseset} (the same set of $u$ reaches $v$), we thus observe: 

\blemma
\label{lemma:labelgenerationcost}
The time complexity of sending vertex label messages $(u,d(u,v))$ along the edges in graph $G$ given an order $\pi$, is the same for PLL and BVC-PLL.  
\elemma 

Following Lemma~\ref{lemma:labelgenerationcost}, we obtain the following corollary. 
\bcoroll
 The total number of distance checks (applying  canonical labeling criterion) being invoked in PLL (Line $5$ in Algorithm~\ref{alg:DLD}) is the same as those being invoked in BVC-PLL (Line $14$ in Algorithm~\ref{alg:BatchedVertexCentricPLL_push}). 
\ecoroll
This is because the number of distance checks is equivalent to the total number of generated distance label message: $\sum_{u \in V } |reach(u)|$ (following the algorithm logic).

\noindent{\bf Cost of Distance Check:}
Now, the cost of the same distance check 
on $d(u,v)$: $d(u,v) < \min_{h \in L(u) \cap L(v)} d(u,h)+d(h,v)$, in PLL and BVC-PLL, is $O(|L(v)|)$. However, $L(v)$ are different for PLL and BVC-PLL:
In PLL, when $u$ reaches $v$, $L(v)$ consists of all vertex labels which have higher rank than $u$; In BVC-PLL, assuming $u$ in batch $B_i$, $L(v)$ consists of all the vertex labels in all the batches before $B_i$ (those are the same as those in PLL) and the vertices in the current batch which are within the distance of $d(u,v)$. 

Given this, let us focus on only those vertices being added at batch $B_i$ for $L(v)$, and denote it as $L^i(v)$. Next, we break the distance check cost on $|L^i(v)|$ into two categories: 1) the {\em positive} distance check which will confirm the vertex $u$ and can add it into the corresponding label of $v$; 2) the {\em negative} distance check will return false on the distance check and thus prune the vertex $u$.  


\bthm (Positive Distance Check) 
\label{thm:positivedistancecheckcost}
The time complexity of all positive distance checks in BVC-PLL is lower than or equal to that of PLL. 
\ethm 
\bproof 
Let us consider any batch $B_i$. 
For the positive cases of distance check $d(u,v)$ here, given a vertex $v$ and $u$, $u$ will always be added to the label of $v$. 
 For PLL, for a vertex $v$, let its complete $\mathcal{L}^i(v)$ consists of $u_1, u_2, \cdots, u_n $ $\in B_i$, where $n=|\mathcal{L}^i(v)|$ and $\pi(u_1)<\pi(u_2) \cdots <\pi(u_n)$. Then the total cost of distance check with respect to $|L^i(v)|$ is simply $0+1+\cdots +n-1 =n(n-1)/2,$ \\
because in PLL, when $u_i$ arrives, $L^i(v)$ already consists of partial labels $\{u_1,\cdots, u_{i-1}\}$.
 For BVC-PLL, for a vertex $v$, we note that its distance label in $B_i$ is arriving in group according to their distances.  Let $g_1,g_2, \cdots g_k$ be the groups ordered by arriving (as well as distance), i.e., given any two vertices $x,y \in g_i$, $d(x,v)=d(y,v)$, and their distance is smaller than those in $g_{i+1}$. Note that {\em for any vertex $u \in g_i$, we only utilize $L^i(v)=g_1 \cup \cdots \cup g_{i-1}$ for distance check} (See Lines $11-15$ in Algorithm~\ref{alg:VertexCentricPLL_push}, $L^i(v)$ will be updated until all the distance checks in a batch $g_i$ are done). Let $n_i=|g_i|$ and $n=\sum_{i=1}^k n_i$, making the total cost of distance checks of vertex $v$ with respect to $|L^i(v)|$ in BVC-PLL to be  

\vspace*{-4.0ex}
\begin{align}
 0+n_1\times n_2+(n_1+n_2)\times n_3 + \cdots + (\sum_{i=1}^{k-1} n_i) \times n_k \nonumber \\
=(n-n_1)n_1+(n-(n_1+n_2))n_2 +\cdots+  (n-\sum_{i=1}^{k-1}n_{k-1}) \nonumber\\ 
=n(n-1)/2-\sum_{i=1}^k n_i(n_i-1)/2. \eproof \nonumber
\end{align}
\vspace*{-2.0ex}

\begin{figure}[t]
    \includegraphics[width=1.0\linewidth]{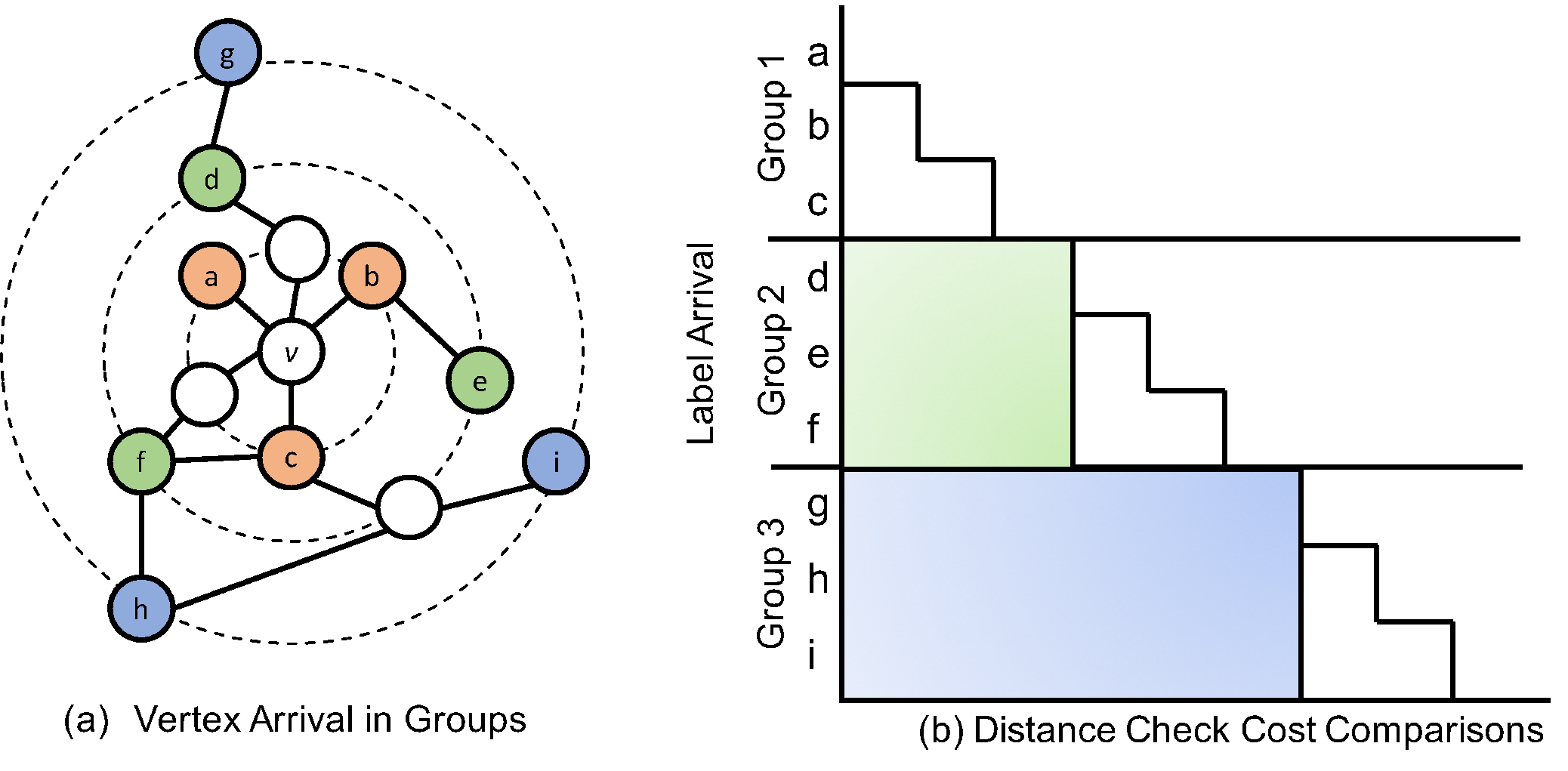}
    \caption{Theorem~\ref{thm:positivedistancecheckcost}}
    \label{figure:positivedistancecheck}
\end{figure}

Figure~\ref{figure:positivedistancecheck} illustrates the key idea in the proof of Theorem~\ref{thm:positivedistancecheckcost}.
Assuming $9$ vertices $a,b,\cdots,i$ in one batch being added into $L(v)$ in PLL labeling, its total distance check cost is $36$ no matter which order they are received in (visualized as the area under the diagonal stairs). Now assuming they arrive in three groups as shown in Figure~\ref{figure:positivedistancecheck}(a), then in BVC-PLL, their total distance check cost is $3 \time 3+3 \times 6=27$, a $25\%$ reduction compared to PLL. 

Theorem~\ref{thm:positivedistancecheckcost} essentially shows that BVC-PLL is able to save the intro-group cross-vertex comparison in each batch. Basically, if vertices arrive at the same time, they have the same distance to vertex $v$ and cannot prune one another.


To compare the time complexity difference between PLL and BVC-PLL for the negative distance check, we introduce the following notation: for any vertex $x$, and one of its vertex label $u$ ($u\in L^i(x)$), we denote $<x,u>$ to be a subset: $\big\{ x \in $
$$\bigcup_{y \in N(x)} L^i(y) \setminus L^i(x): \pi(u) < \pi(v) < \pi(x), d(x,u)> d(v,y)+1 \big\}$$ 
Similarly, we define $<y,v>$ for vertex $y$ with its label $v$, $v\in L^i(y)$: 
$$\big\{ u \in \bigcup_{x \in N(y)} L^i(x) \setminus L^i(y) : \pi(u) < \pi(v), d(x,u)> d(v,y)+1 \big\}$$

\bthm (Negative Distance Check) 
\label{thm:negativedistancecheckcost}
In batch $B_i$, and on negative distance check, the time complexity saved by  BVC-PLL compared with PLL is no higher than  
$$O(\sum_{x\in V} \sum_{u \in L^i(x)} |<x,u>| - \sum_{y \in V} \sum_{v \in L^i(y)} |<y,v>|).$$
The time complexity saved by PLL compared with BVC-PLL is no higher than 
$$O(\sum_{y \in V} \sum_{v \in L^i(y)} |<y,v>|-\sum_{x\in V} \sum_{u \in L^i(x)} |<x,u>|).
$$
\ethm 

Please refer to Appendix for proof.    


Theorem~\ref{thm:negativedistancecheckcost} does not provide a clear winner on the cost of negative check. However, from the symmetric expression of these two qualities, we conjecture they should be close to one another. In Section~\ref{sec:eval}, we will experimentally confirm this. In addition, for negative distance check, we typically do not need to traverse through the entire $L(v)$ set. Indeed, the bit-parallel mechanism proposed in the original PLL paper ~\cite{akiba2013fast} can help provide almost $O(1)$ pruning. Since the number of negative checks is the same for PLL and BVC-PLL, we expect their overall cost will be fairly close to each other. 

\noindent{\bf Putting It  Together:}
Assuming that PLL and BVC-PLL have a similar cost for negative distance checks, theoretically, {\em BVC-PLL may have smaller computational cost than that of PLL (due to positive distance check) since they have the same cost of generating/sending distance labeling!}
Furthermore, BVC-PLL is guaranteed to have a smaller memory access cost for graph topology than PLL as it groups messages together for each edge access. Overall, it seems BVC-PLL, an unexpected marriage between PLL and VC computation, can run faster than the original PLL sequentially and can also enjoy the scalability of the VC model!    
Indeed, Section~\ref{sec:eval} shows that it can be more than two times faster than PLL (both using one thread) on real-world graphs. 





\section{Variants and Implementation}\label{sec:system}


\subsection{Generalization}
\label{subsection:generalization}

\noindent{\bf Directed Graphs:}
For directed graph, each vertex $v$ is assigned with two labels $L_{in}(v)$ and $L_{out}(v)$. VC-PLL and BVC-PLL can be easily extended to handle directed graphs by considering these two labels as separate computations. Specifically, in the Scatter function, the new labels $\delta L_{in}$ and $\delta L_{out}$ will be sent out along the outgoing edges and incoming edges, respectively. In the Gather function, there will be two message queues: one for candidate vertices in $L_{in}$, and another for those in $L_{out}$. The labels generated by this algorithm will be canonical. The computational complexity analysis in Subsections~\ref{subsection:limitations} and ~\ref{subsection:apple2apple} holds for directed graphs as well. 

\noindent{\bf Weighted Graphs:}
The direct application of VC-PLL and BVC-PLL (by changing $d(u,v)+1$ to $d(u,v)+w_e$ where $w_e$ is the edge weight) on weighted graphs can produce a 2-hop labeling; but it may not be a canonical labeling. 
 This is because unlike unweighted graphs, the iteration on the vertex-centric model will not be in sync with the distance between two vertices. For instance, when vertex $u$ reaches $v$ in two iterations, their distance may be larger than a path via vertex $w$ with a higher rank, but $w$ may take more than $2$ iterations to reach $v$ and $u$. Given this, we cannot use the partial label $L(u)$ at an arbitrary iteration to fully determine if vertex $v$ is a true or final label for $u$ anymore. Thus, adding vertex $v$ into $u$'s partial label $L(u)$ or $\delta L(u)$ (using the partial labels in the weighted graph) may lead to unnecessary vertices being spread in the networks.  
 To deal with this problem, at the end of each batch processing (Line $22$ in BVC-PLL), we can perform a distance recheck using only the labels from the batch.
 Since the hash tables of the labeling vertices in the batch are still in the memory, this recheck can be quite efficient.

\subsection{Implementation Issues}
\label{subsection:systemoptimization}

\noindent{\bf Hierarchical Parallelism:} 
The BVC-PLL computation (as shown in Algorithm~\ref{alg:BatchedVertexCentricPLL_push}) is inherently parallel at 
the coarse-grained thread-level. The computation of each batch uses vertex-centric processing (line~\ref{line:vcstart} to line~\ref{line:vcend}) that consists of two parallel phases: ({\em Scatter} and {\em Gather}),  with an implicit synchronization between them. In each phase, each thread processes a chunk of active vertices with dynamic scheduling to achieve load balance. 

In addition, as aforementioned, BVC-PLL is able to significantly increase the data locality for remote vertex memory access, therefore offering us extra opportunities to better exploit fine-grained data-level parallelism (i.e., SIMD parallelism or vectorization).
More specifically, consider the {\em Gather Phase} in Algorithm~\ref{alg:BatchedVertexCentricPLL_push} that involves an intensive label distance check kernel (line~\ref{line:lcstart} to line~\ref{line:lcend}). BVC-PLL can vectorize this kernel with the help of advanced SIMD gather/scatter and mask instructions in the latest AVX512 intrinsic set\footnote{\url{https://software.intel.com/sites/landingpage/IntrinsicsGuide/}}.

Moreover, for weighted graphs, the distance recheck operation incurs extra overheads. The hierarchical parallelism is also applied to address this challenge. In particular,  efficient SIMD parallelism significantly reduces the overhead of distance rechecks.   

\noindent{\bf Integrated Bitmap and Queue:}
Much temporary data is 
generated for both labeling vertices and active vertices during each batch processing.  
These steps require a {\em clearance} (e.g., Algorithm~\ref{alg:BatchedVertexCentricPLL_push}, line~\ref{line:clear}). The cost of this clearance is significant as this operation occurs  for each batch. 
Traditionally, we often use  either a bitmap or a queue to handle the set of active vertices. However, they become inefficient or insufficient for supporting BVC-PLL. For a bitmap, each of its cleanings can take $O(|V|)$ where $|V|$ is the total number of vertices; for a queue, it cannot support efficient checks for whether 
a given vertex is active or not. Given this, we propose a new traversal control data structure by combining both the bitmap and the queue. 
The basic idea is that a bitmap supports fast recording and checking visited vertices and a queue supports fast finding and clearing the visited vertices. Each time a vertex is processed, we add it to both the bitmap and the queue.
This approach is different from the bitmap and queue used in the push and pull strategy presented in ~\cite{patterson2012direction,shun2013ligra,besta2017push} because we use both the bitmap and queue simultaneously rather than in different stages of processing.  

\noindent{\bf Bit-parallel Adoption:}
Similar to PLL~\cite{akiba2013fast}, {\em bit-parallel} is also adopted to accelerate the {\em distance checking} in the implementation of BVC-PLL for unweighted graphs. Its construction is similar to multi-source BFS traversals and can be easily expressed in the Vertex-Centric computing model. 



\section{Evaluation}\label{sec:eval}
In this section, we perform a detailed evaluation of BVC-PLL,  
focusing on answering  the following questions: 1) How does BVC-PLL algorithm perform against the original PLL  
in a sequential setting (single thread; no parallelism)? Specifically, the theoretical analysis indicates it may run faster, but we conduct experiments to confirm this.  2) How does BVC-PLL scale as the number of threads increases? 3) The breakdown of  runtime of BVC-PLL, and more specifically, how does the  theoretical cost analysis align with 
experimentation on real-world graphs, such as positive and negative distance checks and memory access for graph topology? 
4) How does the weighted extension of BVC-PLL perform and how does it fare against ParaPLL~\cite{qiu2018parapll} (the state-of-the-art parallel weighted PLL algorithm)? 



\subsection{Experimental Setup}

\noindent{\bf Platform:} We perform all the experiments on an Intel Xeon Gold 6138 CPU. It is a Skylake processor with 20 cores running at 2.0 GHz supporting efficient 512-bit AVX-512 intrinsics,
with 27.5 MB L3 cache and 192 GB DDR4 memory shared among all cores. All code is compiled with an Intel icc compiler (version 19.0.2.187) with {\tt -O3} optimization option. 
Hyper-threading is not used to simplify the analysis of experiment results.
   
\noindent{\bf Graph Datasets:} The 10 graphs used in our evaluation are characterized in Table~\ref{tab:datasets}. They are from 5 categories (Social, Citation, Communication, Hyperlink, and Computer) with varied numbers of vertices, edges, degrees, and diameters --  {\tt GNUTELLA}, and {\tt WIKITALK} are from SNAP\footnote{\url{https://snap.stanford.edu/snap/}}, 
{\tt DBLP}, {\tt YOUTUBE}, {\tt TREC WT10G}, {\tt SKITTER}, {\tt CATS-DOGS}, and {\tt FLICKR} are from KONET\footnote{\url{http://konect.uni-koblenz.de/}}, and
{\tt HOLLYWOOD} and {\tt INDOCHINA} are from SuiteSparse Matrix Collection\footnote{\url{https://sparse.tamu.edu/}}.
Particularly, {\tt HOLLYWOOD} and {\tt INDOCHINA} are large graphs with 100 - 200 M edges.
These graphs are all unweighted.  
To test the performance of our BVC-PLL on weighted graphs, we randomly assign weights (from 1 to 7  with a  uniform distribution) to their edges.
Since we only evaluate  algorithms for undirected graphs, we have transformed the edges in the directed graphs in Citation and Hyperlink as undirected edges.  

\begin{table}[]
\small
\caption{List of datasets. ``deg.'' denotes average degree. ``dia.'' denotes diameter. 
}
\label{tab:datasets}
\begin{tabular}{|c||c|r|r|c|c|}
\hline
Graph      & Category  & \multicolumn{1}{c|}{$|V|$} & \multicolumn{1}{c|}{$|E|$} & deg. & dia. \\ 
\hline
\hline
GNUTELLA   & Social    & 63 K                     & 148 K                    & 4.7    & 11     \\
DBLP       & Citation  & 317 K                    & 1 M                      & 7.9    & 10     \\
WIKITALK   & Comm.     & 2.4 M                    & 5 M                      & 2.1    & 9      \\
YOUTUBE    & Social    & 3.2 M                    & 9 M                      & 5.8    & 31     \\
TREC WT10G & Hyperlink & 1.6 M                    & 8.0 M                    & 10.1   & 112    \\
SKITTER    & Computer  & 1.7 M                    & 11 M                     & 13.1   & 31     \\
CATS-DOGS  & Social    & 62 K                     & 15 M                     & 50.3   & 15     \\
FLICKR     & Social    & 2.3 M                    & 33 M                     & 28.8   & 23     \\
HOLLYWOOD  & Social    & 1.1 M                    & 114 M                    & 99     & 9      \\
INDOCHINA  & Hyperlink & 7.4 M                    & 194 M                    & 52     & 207    \\ 
\hline
\end{tabular}
\end{table}

\noindent{\bf Benchmarks:}
For the sequential performance comparison on unweighted graphs, we compare BVC-PLL against the PLL implementations by the original authors~\cite{akiba2013fast}, and by ~\cite{li2017experimental}. We found these two implementations provide comparable performance with the former being slightly faster. Given this, we only report its PLL performance result below. 
For the scalability performance comparison on weighted graphs, we compare the weighted BVC-PLL against the implementation of ParaPLL~\cite{qiu2018parapll}.   
For the vertex order, we adopt the original and the most popular method where  the vertices are ordered by their vertex degree~\cite{akiba2013fast,li2017experimental}. 

\noindent{\bf Batch Size of BVC-PLL:}
Throughout the experiments, we use $1024$ as the batch size for unweighted graph and use $512$ for weighted graph. 
In general, we observe the larger the batch size the better performance if the memory can afford such batch size. In our experimental platform, we found those two are the optimal batch size. Due to the space limitation, we will not report the performance results with respect to batch size below.

\subsection{BVC-PLL vs PLL and Scalability}\label{sec:eval:subsec:perf}

\begin{table}[]
\small
\caption{{\bf Performance}: BVC-PLL vs. PLL}  \footnote{\scriptsize LT denotes labeling time (s). $|L|$ denotes average label size for each vertex. SP denotes speedup. BVC-PLL and PLL has the same label size. This is the same to Table~\ref{tab:weighted_performance}}

\label{tab:unweighted_performance}
\begin{tabular}{|c|r||r|rr|}
\hline
\multirow{2}{*}{Graph} & \multicolumn{1}{c||}{\multirow{2}{*}{$|L|$}} & \multicolumn{1}{c|}{PLL}    & \multicolumn{2}{c|}{BVC-PLL}                            \\
                         & \multicolumn{1}{c||}{}                    & \multicolumn{1}{c|}{LT} & \multicolumn{1}{c}{LT} & \multicolumn{1}{c|}{SP} \\ 
\hline
\hline
GNUTELLA                 & 477                                      & 33                        & \textbf{13}              & 2.46                    \\
DBLP                     & 214                                      & 61                        & \textbf{47}              & 1.30                    \\
WIKITALK                 & 12                                       & 40                        & \textbf{32}              & 1.24                    \\
YOUTUBE                  & 70                                       & 285                       & \textbf{249}             & 1.15                    \\
TREC WT10G                & 269                                      & 462                       & \textbf{323}             & 1.43                    \\
SKITTER                  & 138                                      & 317                       & \textbf{242}             & 1.31                    \\
CATS-DOGS                   & 96                                       & 117                       & \textbf{92}              & 1.28                    \\
FLICKR                   & 442                                      & 1,624                     & \textbf{909}             & 1.79                    \\
HOLLYWOOD                & 2,199                                    & 10,743                    & \textbf{4,368}           & 2.46                    \\
INDOCHINA                & 442                                      & 4,755                     & \textbf{3,508}           & 1.36                    \\ 
\hline
\end{tabular}
\end{table}

Table~\ref{tab:unweighted_performance} shows the performance comparison between  BVC-PLL as a sequential algorithm and PLL (both using single thread and no other parallelism, such as SIMD) on all graphs. Both algorithms use the same vertex order and produce the same label size,  as expected. Interestingly, the BVC-PLL algorithm consistently outperforms PLL with the speedup ranging from 1.15X ({\tt YOUTUBE}) to 2.46X ({\tt GNUTELLA} and {\tt HOLLYWOOD}) with an average speedup 1.58X 
over PLL. This observation is consistent with our theoretical analysis in Subsection~\ref{subsection:apple2apple}. 
In the next subsection, we will perform a more detailed cost breakdown and comparison. 


\begin{figure}[t]
    \centering
    \begin{subfigure}[t]{0.4\textwidth}
        \includegraphics[width=\textwidth]{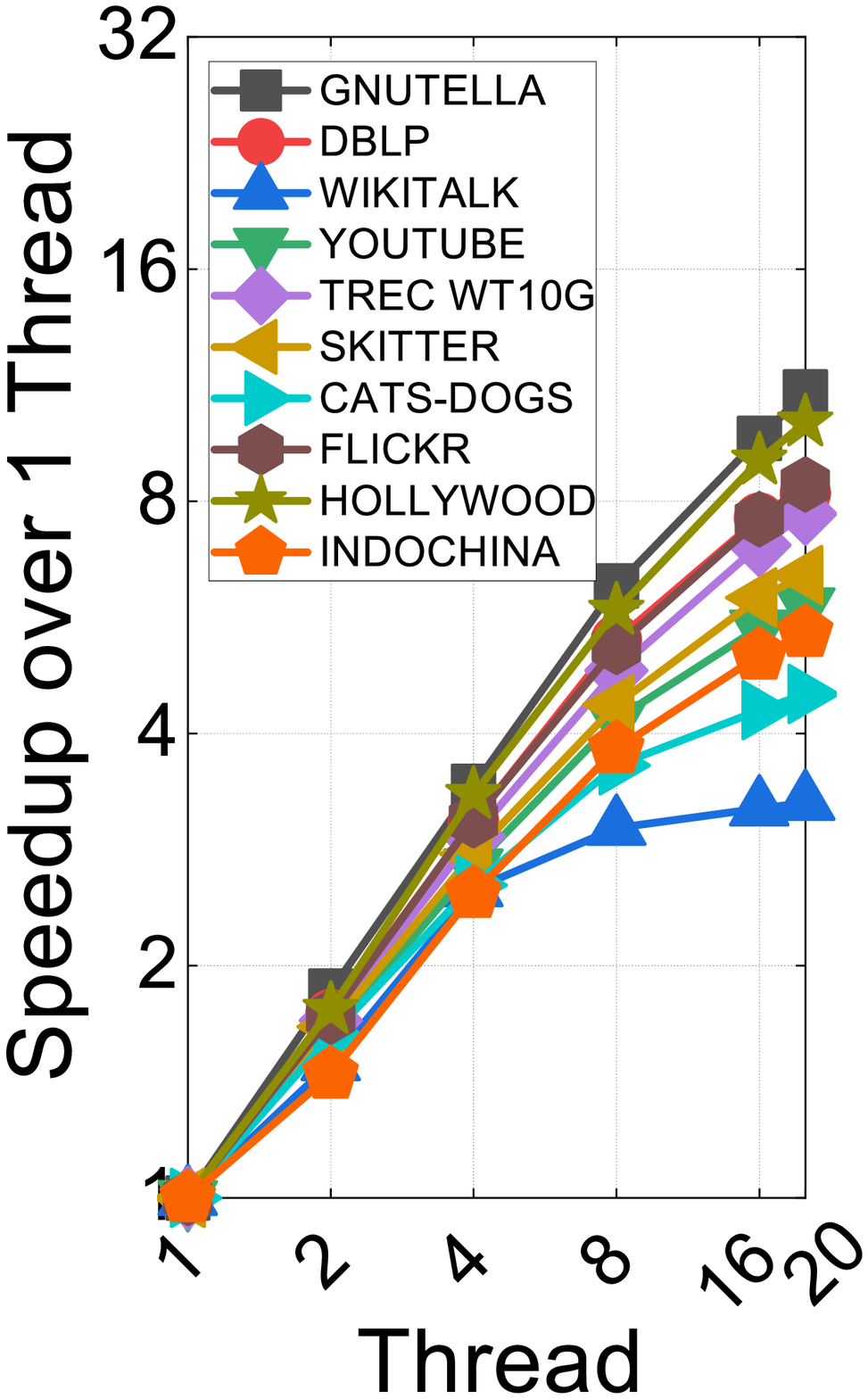}
        \caption{Speedup over 1 thread} 
        \label{fig:6_pado_unw_unv_para_scal}
    \end{subfigure}
    \hfill
    \begin{subfigure}[t]{0.4\textwidth}
        \includegraphics[width=\textwidth]{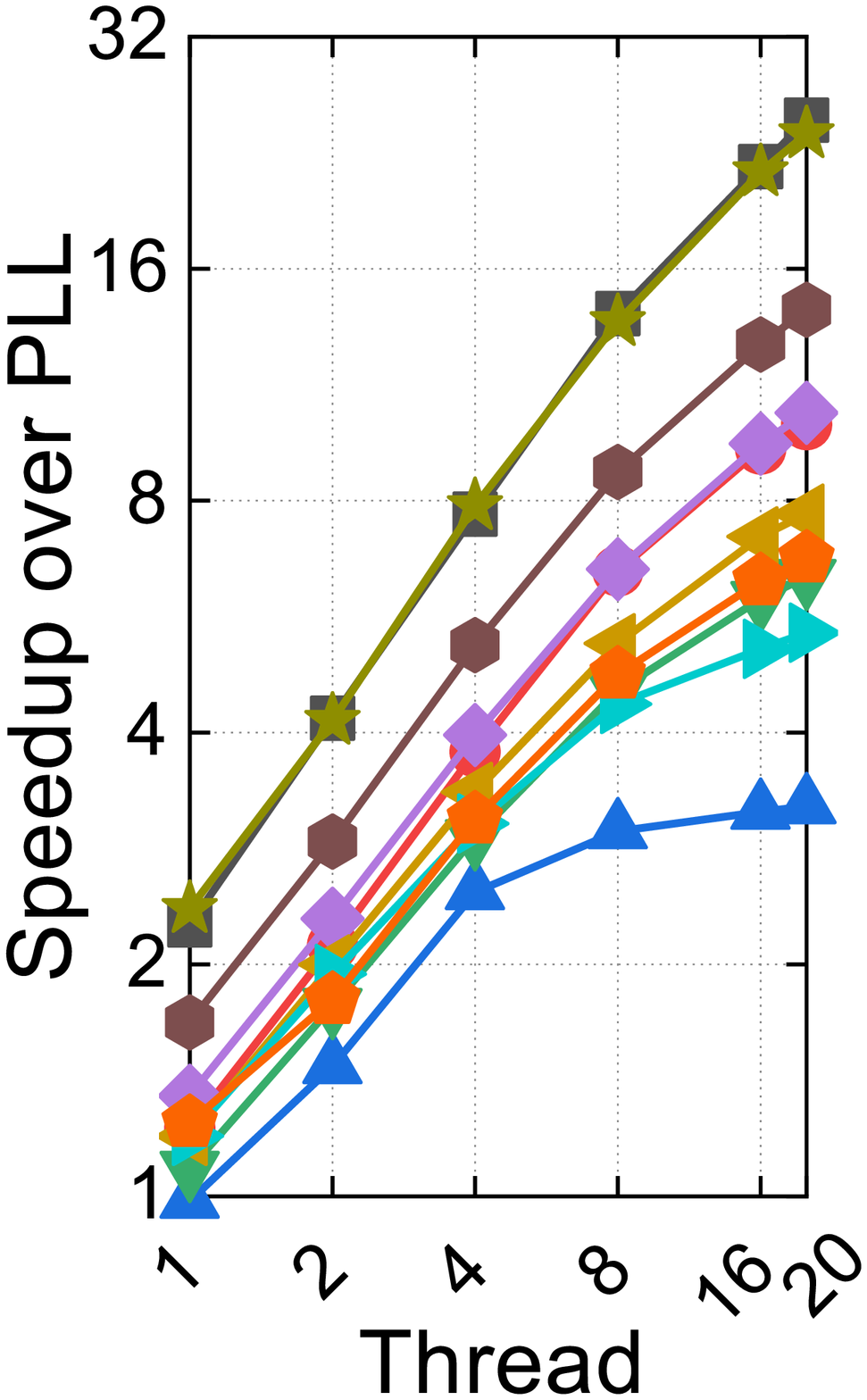}
        \caption{Speedup over PLL} 
        \label{fig:6_pado_over_pll_unw_unv_unp_scal}
    \end{subfigure}
    \caption{The scalability of BVC-PLL (unweighted)}
    \label{fig:unweighted_scalability}
\end{figure}

Figure~\ref{fig:unweighted_scalability} shows the scalability of BVC-PLL on all graphs.  Figure~\ref{fig:6_pado_unw_unv_para_scal} shows its speedup over 1-thread BVC-PLL, while Figure~\ref{fig:6_pado_over_pll_unw_unv_unp_scal} shows its speedup over the original sequential PLL.
With $20$ threads, BVC-PLL can achieve up to $14.71$ and $33.11$ speedup over its 1-thread version and PLL, respectively,  demonstrating good scalability. 

In addition, by comparing Figure~\ref{fig:unweighted_scalability} and the average label size of each vertex in Table~\ref{tab:unweighted_performance}, we found that generally, BVC-PLL scales better as the average label size increasing. 
For example, {\tt GNUTELLA} and {\tt HOLLYWOOD} with the largest average label sizes result in the best scalability while {\tt WIKITALK} with the smallest results in the worst scalability. The labeling size provides a good indication of the total computational costs (message passing and distance checks) involved for each vertex. The better scalability of larger labeling sizes is consistent with the computing scalability of multi-core architecture.


\subsection{Understanding the Performance}\label{sec:underlying}

\begin{figure}[t]
    \centering
    \begin{subfigure}[t]{0.4\textwidth}
  \includegraphics[width=1.05\textwidth]{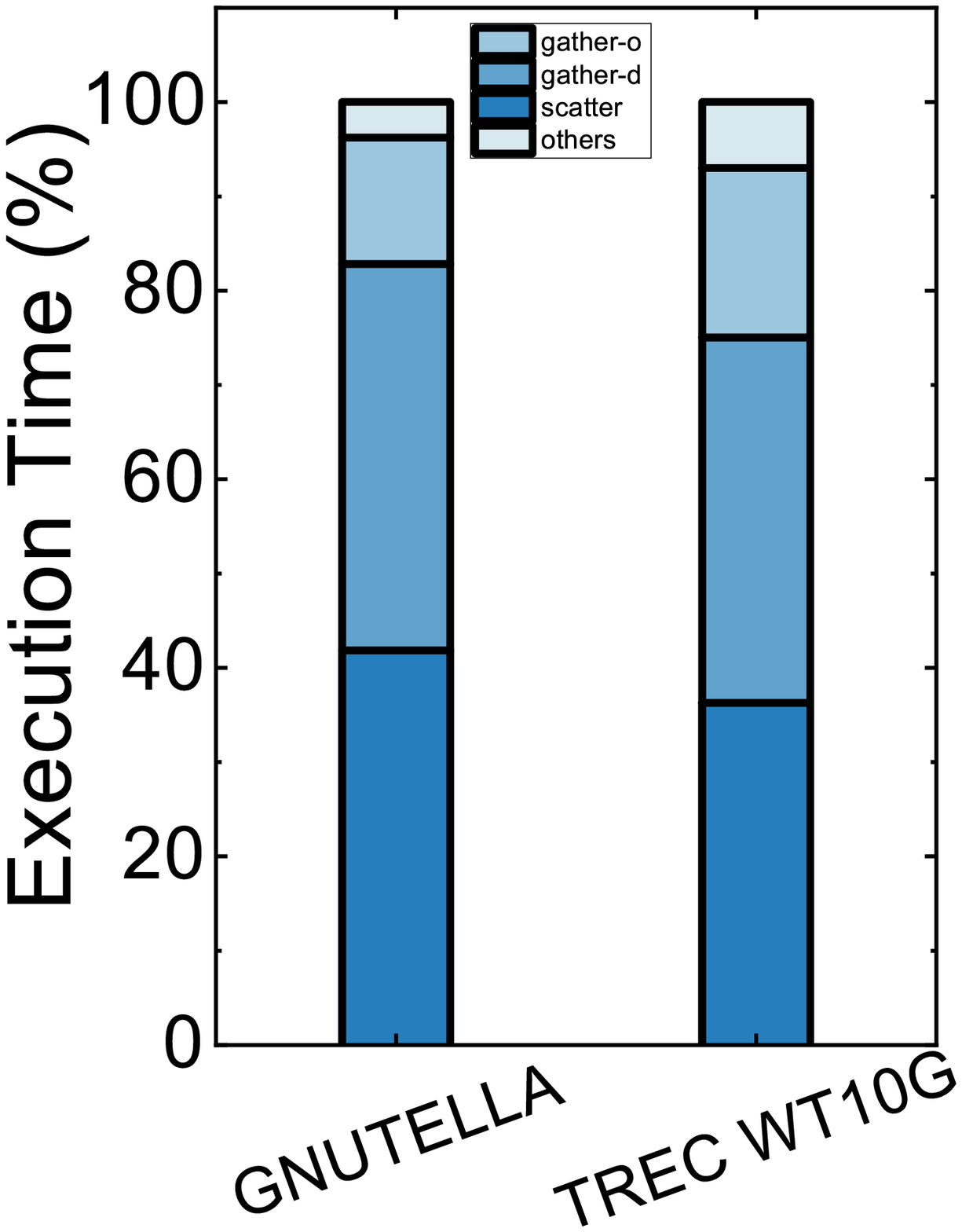}
\caption{BVC-PLL execution time breakdown.} 
\label{fig:6_pado_exe_time_breakdown}
    \end{subfigure}
    \hfill
    \begin{subfigure}[t]{0.4\textwidth}
\includegraphics[width=\textwidth]{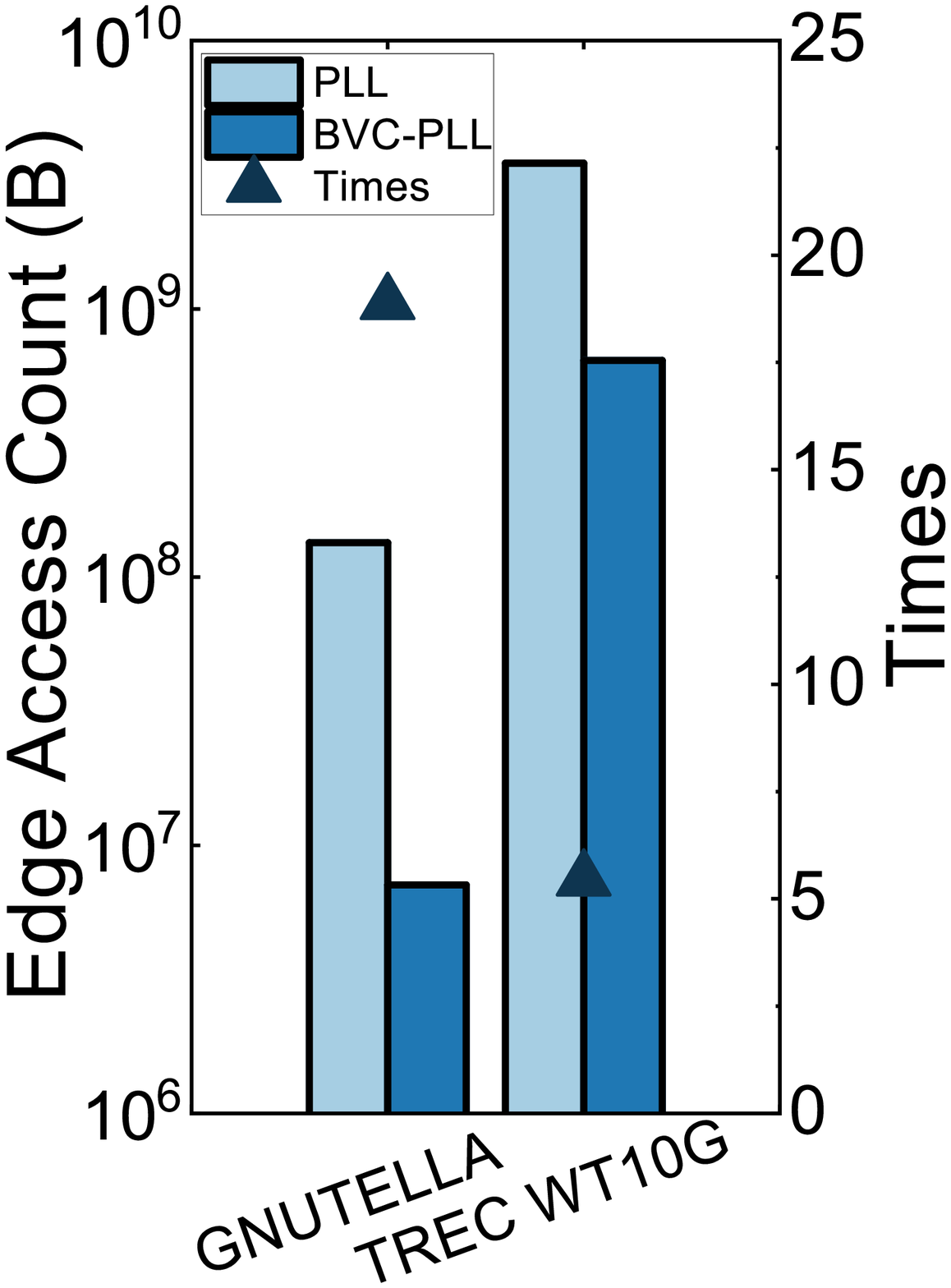}
\caption{Edge access count: BVC-PLL vs. PLL.} 
\label{fig:6_pado_vs_pll_edge_access_count}
    \end{subfigure}
    \caption{Performance Analysis}
    \label{fig:performanceanalysis1}
\end{figure}

Figure~\ref{fig:6_pado_exe_time_breakdown} shows the overall running time breakdown on two graphs: GNUTELLA and TREC WT10G. Due to space limitation, we only report two -- trends are similar in other graphs.   We can see the Gather  and the Scatter phases dominate the overall computational costs. In addition, within gather, the distance check time takes about $60\%-80\%$  and $30\%-40\%$ of the gather phase and overall time, respectively.    

Table ~\ref{tab:dist_check_Lv_size_pado_vs_pll} shows the theoretical computational distance check cost, $\sum L(v)$, as being defined in Subsection~\ref{subsection:apple2apple}. We can see that the total cost of the positive distance checks from BVC-PLL is strictly smaller than that from PLL (ranging from $1.03$ to $2$), on average $1.23$ times smaller.
Also, the theoretical negative distance check cost is indeed very close to each other, thus experimentally confirming our conjecture. 




\begin{table}[t]
\scriptsize
\caption{The sum of the size of $L(v)$: BVC-PLL vs. PLL. ``pos.'' denotes positive case. ``neg.'' denotes negative case. Unit (Billion)}
\label{tab:dist_check_Lv_size_pado_vs_pll}
\begin{tabular}{|c|rr|rr|rr|}
\hline
\multirow{2}{*}{Graph} & \multicolumn{2}{c|}{PLL}                                     & \multicolumn{2}{c|}{BVC-PLL}                                 & \multicolumn{2}{c|}{Ratio}                           \\
                       & \multicolumn{1}{c}{pos.} & \multicolumn{1}{c|}{neg. } & \multicolumn{1}{c}{pos.} & \multicolumn{1}{c|}{neg.} & \multicolumn{1}{c}{pos.} & \multicolumn{1}{c|}{neg.} \\ \hline
GNUTELLA               & 22                           & \textbf{23}                   & \textbf{19}                  & 24                            & 1.18                     & 0.96                      \\
DBLP                   & 46                           & \textbf{36}                   & \textbf{41}                  & 38                            & 1.14                     & 0.96                      \\
WIKITALK               & 32                           & 19                            & \textbf{16}                  & \textbf{19}                   & 2.00                     & 1.00                      \\
YOUTUBE                & 108                          & 138                           & \textbf{80}                  & \textbf{138}                  & 1.35                     & 1.00                      \\
TREC WT10G             & 314                          & \textbf{103}                  & \textbf{300}                 & 104                           & 1.05                     & 0.99                      \\
SKITTER                & 260                          & 513                           & \textbf{208}                 & \textbf{508}                  & 1.25                     & 1.01                      \\
CATS-DOGS              & 44                           & 244                           & \textbf{37}                  & \textbf{244}                  & 1.19                     & 1.00                      \\
FLICKR                 & 893                          & \textbf{2,003}                & \textbf{830}                 & 2,028                         & 1.08                     & 0.99                      \\
HOLLYWOOD              & 6,667                        & \textbf{38,343}               & \textbf{6,455}               & 38,487                        & 1.03                     & 1.00                      \\
INDOCHINA              & 6,265                        & \textbf{1,886}                & \textbf{5,973}               & 1,899                         & 1.05                     & 0.99                      \\ \hline
\end{tabular}
\end{table}

\begin{figure}[t]
    \centering
    \begin{subfigure}[t]{0.4\textwidth}
        \includegraphics[width=\textwidth]{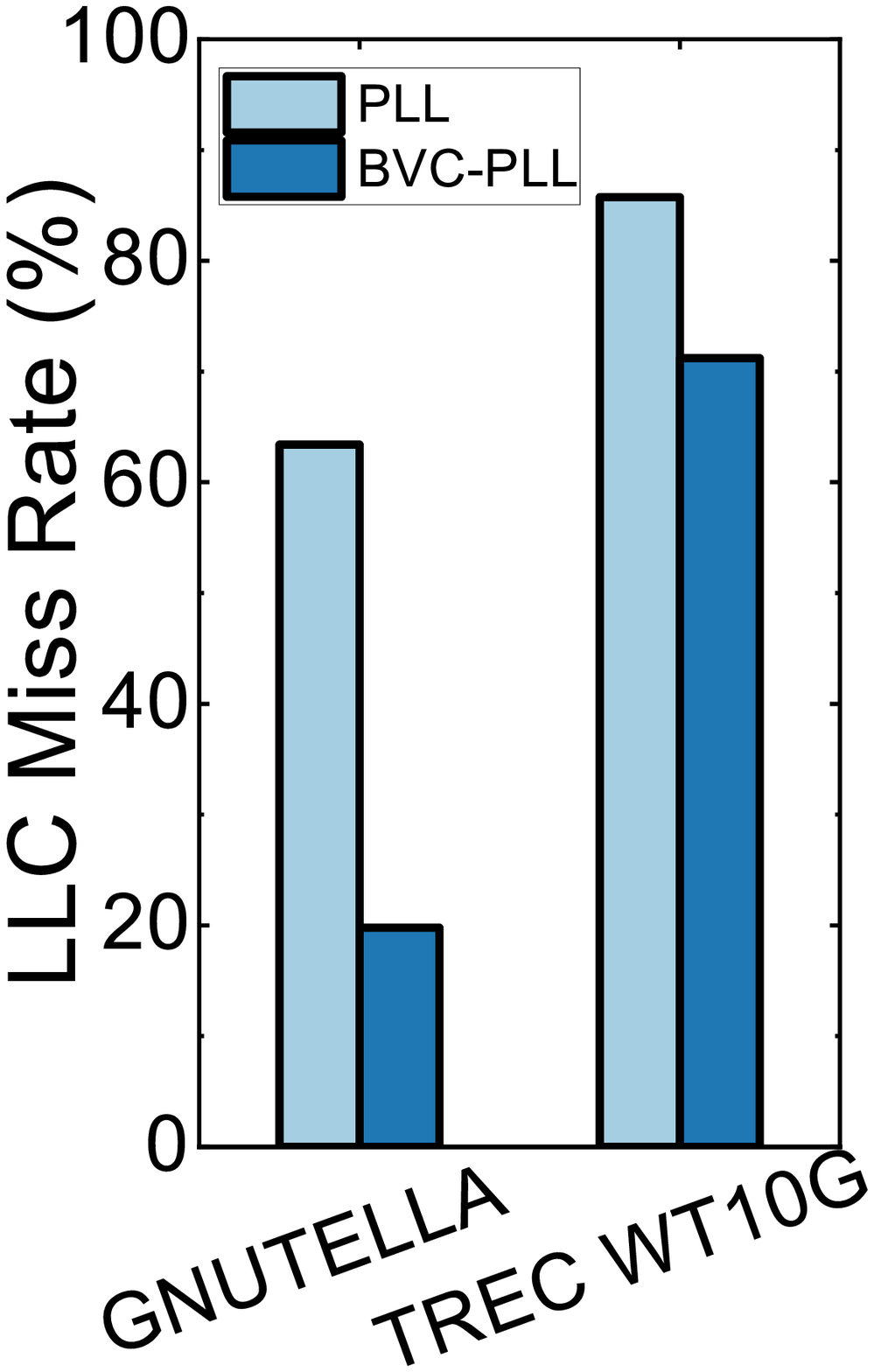}
        \caption{LLC miss rate} 
        \label{fig:6_pado_vs_pll_llc_miss_rate}
    \end{subfigure}
    \hfill
    \begin{subfigure}[t]{0.4\textwidth}
        \includegraphics[width=\textwidth]{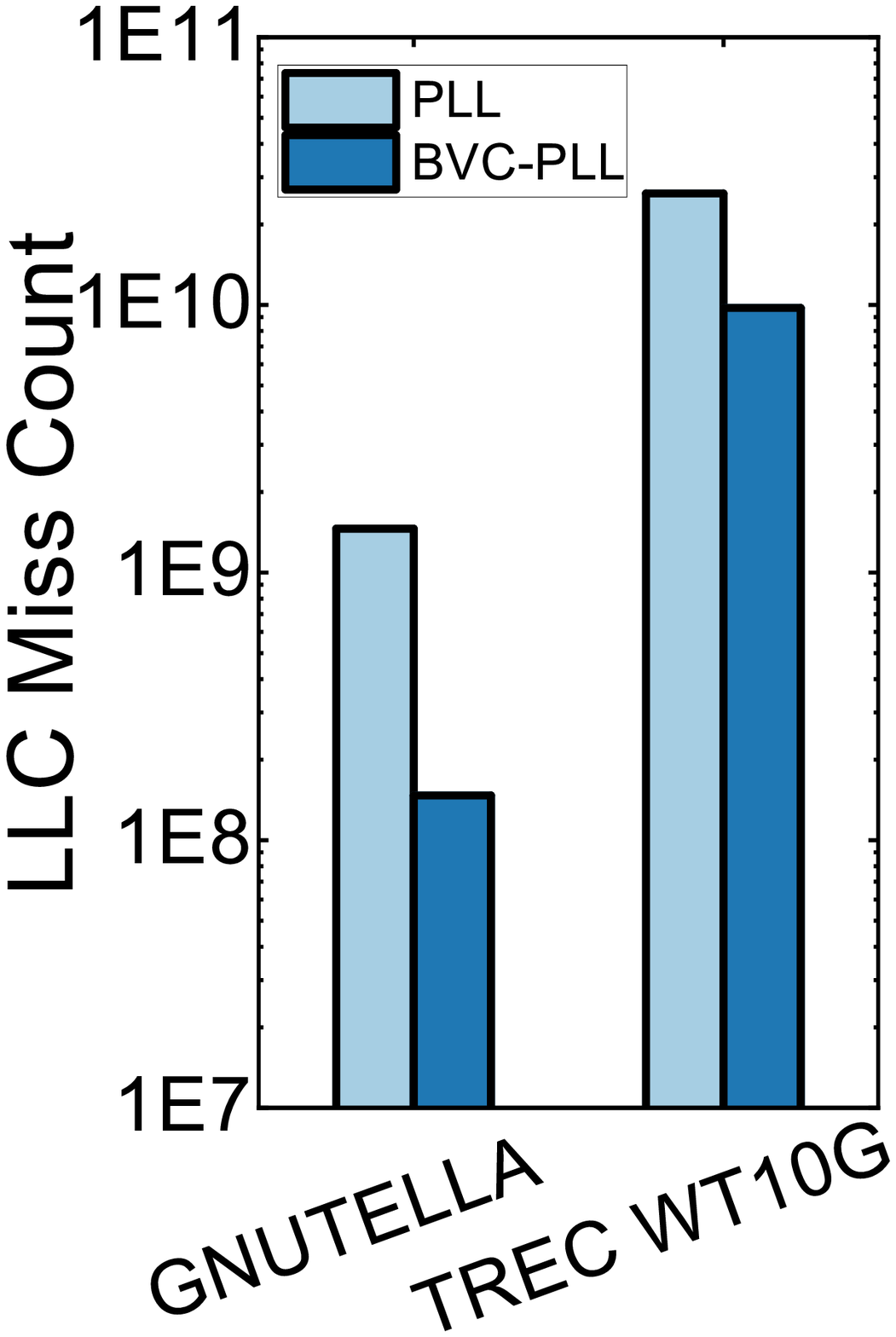}
        \caption{LLC miss count} 
        \label{fig:6_pado_vs_pll_llc_miss_count}
    \end{subfigure}
    \caption{Data locality: BVC-PLL vs. PLL}
    \label{fig:data_locality_pado_vs_pll}
\end{figure}

Figure~\ref{fig:6_pado_vs_pll_edge_access_count} shows the total number of edge access for PVC-PLL and PLL on two graphs: GNUTELLA and TREC WT10G. We can see that PVC-PLL has $5$ and $18$ times reduction for these two graphs! This also confirms our theoretical analysis on the reduced memory access for graph topology.
Finally, Figure~\ref{fig:data_locality_pado_vs_pll} shows the LLC (last level cache) miss rate and miss access count for the whole labeling process of BVC-PLL and PLL.
Again, we can see BVC-PLL has consistent lower LLC miss rate and access count than PLL!







\subsection{Extension  to Weighted Graphs}

\begin{table}[t]
\small
\caption{{\bf Weighted Performance}: BVC-PLL vs. PLL (Dijkstra). ``-S'' denotes SIMD version. ``-N'' denotes non-SIMD version.}
\label{tab:weighted_performance}
\begin{tabular}{|@{}c@{}|@{}r@{}||@{}rr@{}|@{}rr@{}@{}r@{}@{}r@{}|}
\hline
\multirow{2}{*}{Graph} & \multicolumn{1}{@{}c@{}||}{\multirow{2}{*}{$|L|$}} & \multicolumn{2}{@{}c@{}|}{PLL}                                               & \multicolumn{4}{@{}c@{}|}{BVC-PLL}                                                                                                           \\
                       & \multicolumn{1}{@{}c@{}||}{}                    & \multicolumn{1}{@{}c@{}}{LT-N} & \multicolumn{1}{@{}c@{}|}{LT-S} & \multicolumn{1}{@{}c@{}}{LT-N} & \multicolumn{1}{c@{}}{LT-S} & \multicolumn{1}{@{}c@{}}{SP-N} & \multicolumn{1}{c@{}|}{SP-S} \\ 
\hline
\hline
GNUTELLA               & 656                                      & 52                                & 47                             & 123                               & \textbf{37}                   & 0.42                            & 1.26                         \\
DBLP                   & 387                                      & 152                               & \textbf{139}                   & 293                               & 146                           & 0.52                            & 0.95                         \\
WIKITALK               & 152                                      & 350                               & 327                            & 396                               & \textbf{171}                  & 0.88                            & 1.92                         \\
YOUTUBE                & 147                                      & 652                               & 625                            & 763                               & \textbf{546}                  & 0.85                            & 1.14                         \\
TREC WT10G             & 304                                      & 632                               & \textbf{579}                   & 1,140                             & 662                           & 0.55                            & 0.87                         \\
SKITTER                & 432                                      & 1,511                             & 1,467                          & 2,146                             & \textbf{921}                  & 0.70                            & 1.59                         \\
CATS-DOGS              & 224                                      & 527                               & 510                            & 442                               & \textbf{347}                  & 1.19                            & 1.47                         \\
FLICKR                 & 653                                      & 3,879                             & 3,826                          & 4,189                             & \textbf{2,483}                & 0.93                            & 1.54                         \\
HOLLYWOOD              & 2,217                                    & 18,707                            & 18,041                         & 24,161                            & \textbf{10,399}               & 0.77                            & 1.73                         \\
INDOCHINA              & 828                                      & 13,940                            & \textbf{13,105}                & 26,744                            & 14,768                        & 0.52                            & 0.89                         \\
\hline
\end{tabular}
\end{table}


\begin{figure}[t]
    \centering
    \begin{subfigure}[t]{0.4\textwidth}
        \includegraphics[width=\textwidth]{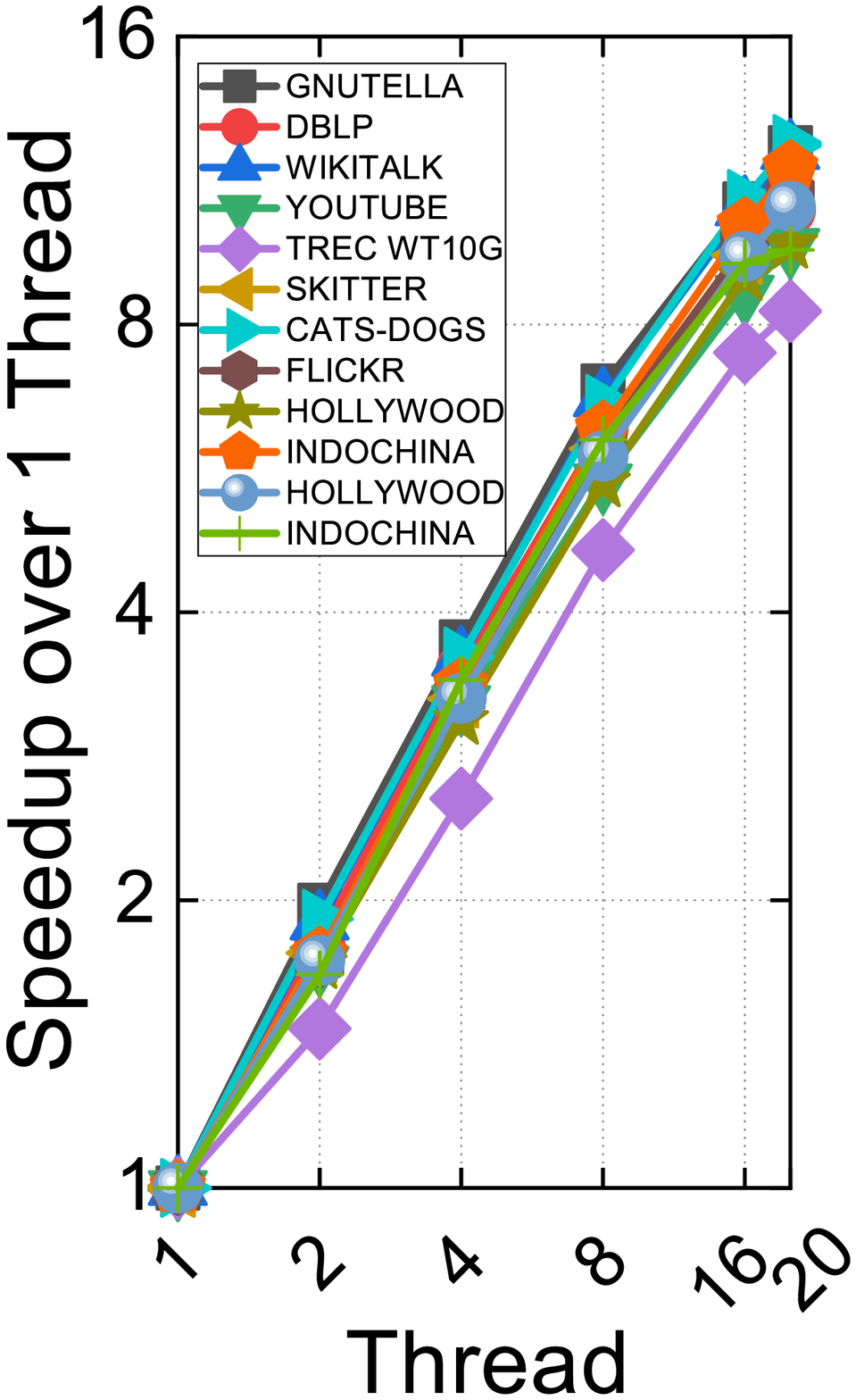}
        \caption{Speedup over 1 thread} 
        \label{fig:6_pado_weighted_v_p_scalability}
    \end{subfigure}
    \hfill
    \begin{subfigure}[t]{0.4\textwidth}
        \includegraphics[width=\textwidth]{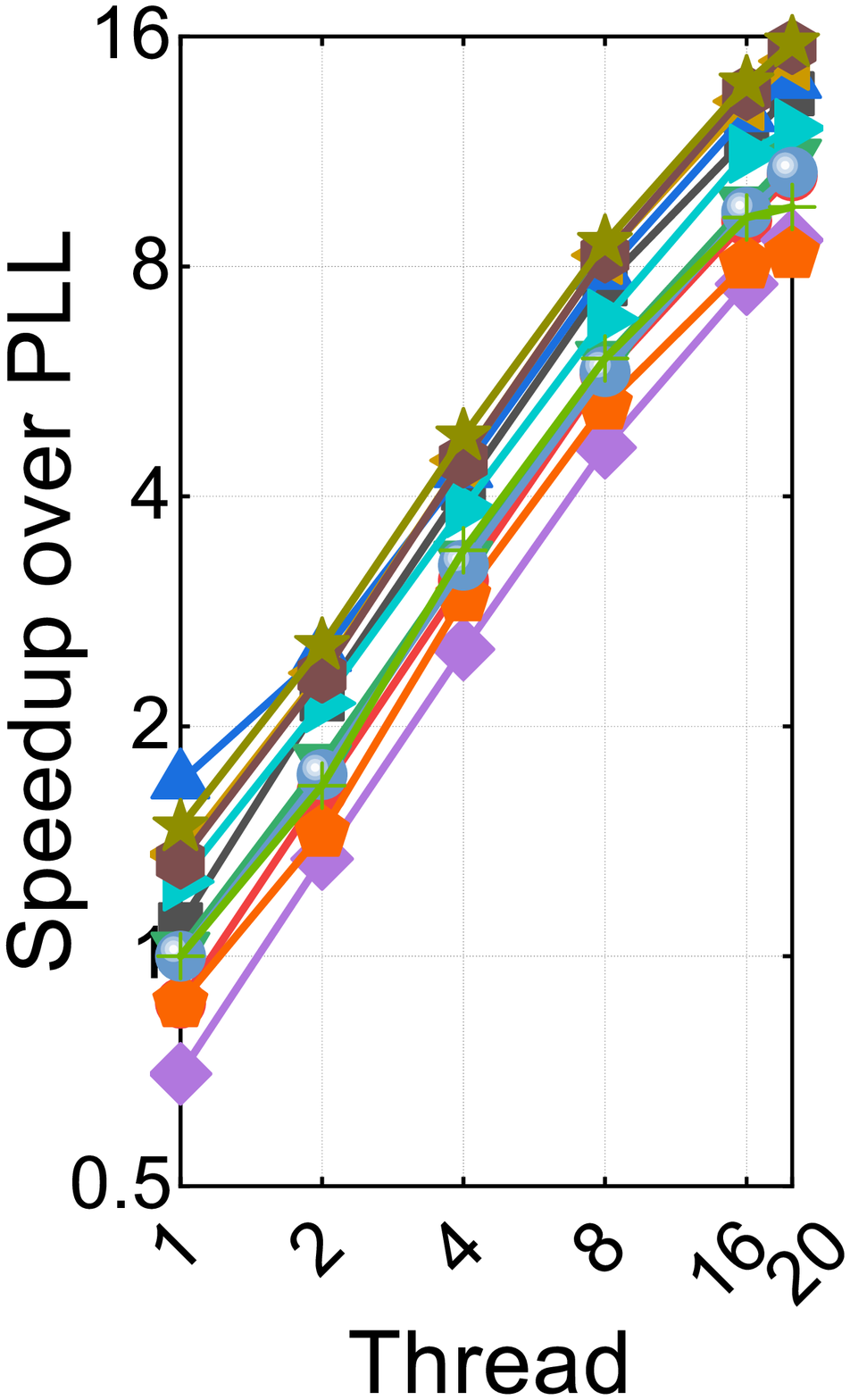}
        \caption{Speedup over PLL (Dijkstra)} 
        \label{fig:6_pado_weighted_over_pll_v_p_scalability}
    \end{subfigure}
    \caption{The scalability of BVC-PLL (weighted)}
    \label{fig:weighted_scalability}
\end{figure}

\begin{figure}[t]
\centering
\includegraphics[width=0.8\textwidth]{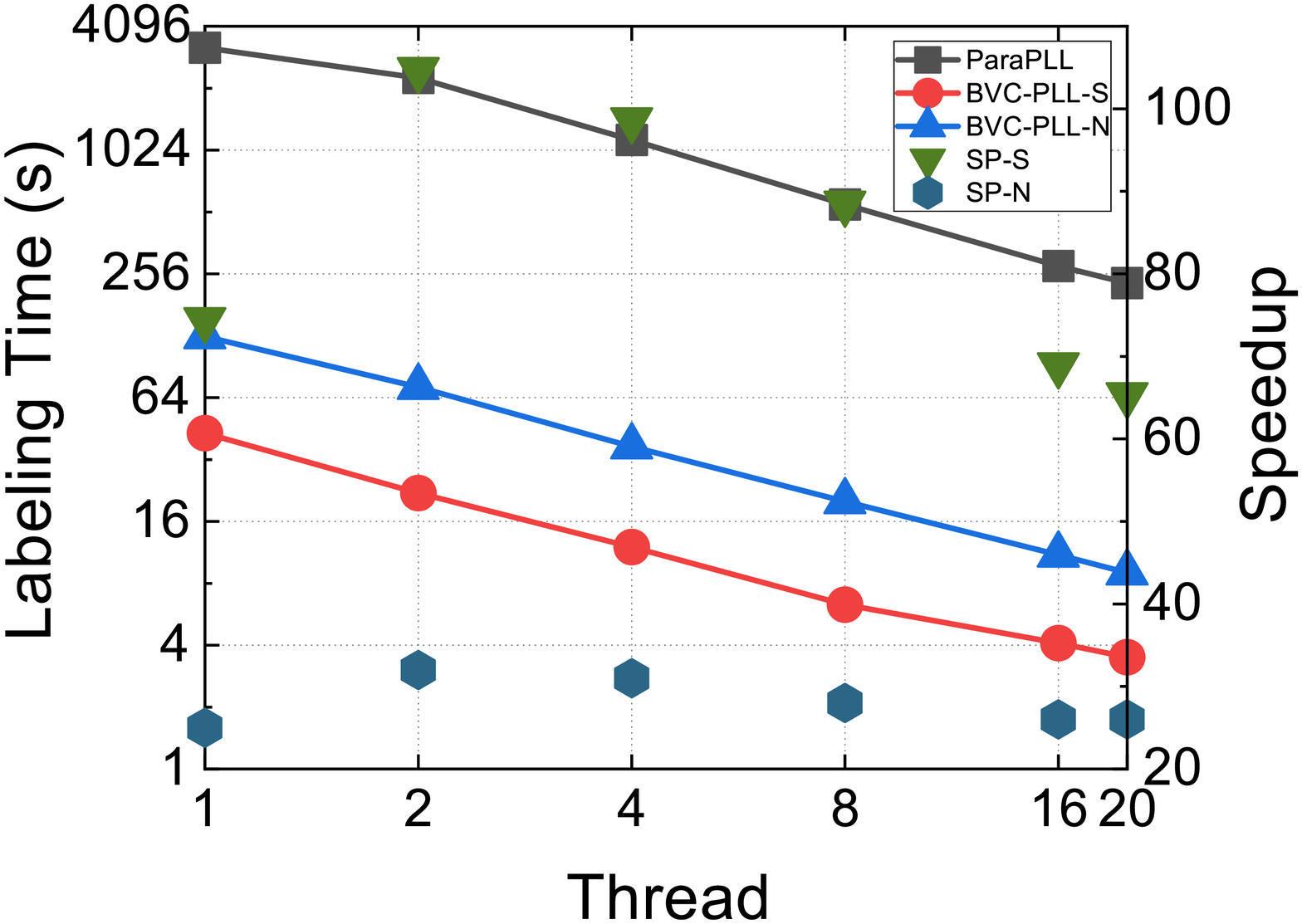}
\caption{{\bf Parallel Weighted Performance}: BVC-PLL vs. ParaPLL on GNUTELLA.} 
\label{fig:6_pado_vs_parapll_performance}
\end{figure}

A similar performance study is conducted between PLL and BVC-PLL for weighted graphs.
To evaluate weighted  BVC-PLL's sequential performance against PLL, we have modified the original PLL implementation as suggested in ~\cite{akiba2013fast}, changing its BFS traversal to Djkstra's algorithm.  We also  extended BVC-PLL as described in  Subsection~\ref{subsection:generalization}. 
Please notice: both PLL and BVC-PLL are optimized with SIMD for the weighted version (and for the unweighted version, we also implemented them with SIMD however without obvious speedup change).


Table~\ref{tab:weighted_performance} shows the comparison results for 1-thread SIMD and non-SIMD versions of PLL and BVC-PLL.  For all non-SIMD tests, PLL consistently performs better than BVC-PLL; while for most SIMD tests, BVC-PLL outperforms PLL. This is because 
the weighted BVC-PLL introduces additional distance check (due to additional message passing) and rechecks, which  significantly increases the number of instructions for BVC-PLL, resulting in degraded performance.  However, SIMD parallelism is a good remedy that can significantly reduce the number of instructions. 
It should be noted that BVC-PLL is able to effectively exploit SIMD parallelism   because  the data locality 
has been improved. (See the performance analysis in last Subsection).  
In particular, for SIMD version, BVC-PLL outperforms PLL for 7 out 10 graphs, resulting in 1.14X to 1.92X speedup with an average of 1.34X.  
For the slowdown cases, BVC-PLL's performance is only degraded up to around $10\%$. Please notice that our BVC-PLL is able to continue exploring hierarchical parallelism to further extract the most out of the massive parallelism of modern processors.

Figure~\ref{fig:weighted_scalability} shows the scalability of BVC-PLL on all weighted graphs, in which, Figure~\ref{fig:6_pado_weighted_v_p_scalability} shows its speedup over 1-thread BVC-PLL while Figure~\ref{fig:6_pado_weighted_over_pll_v_p_scalability} shows its speedup over PLL. With 20 threads, BVC-PLL can achieve up to 13X and 16X speedup over its 1-thread version and PLL, demonstrating good scalability.

Finally, we compare BVC-PLL with the state-of-the-art ParaPLL, which does  weighted parallel PLL. 
Unfortunately, it can only run on small graphs (this is consistent on what being presented in their original paper~\cite{qiu2018parapll}). In Figure~\ref{fig:6_pado_vs_parapll_performance} shows the performance comparison of BVC-PLL and ParaPLL on the graph GNUTELLA (the only graph we are able to run for ParaPLL, as it throws an error of {\tt Segmentation Fault} with the other graphs). For this graph, we can see that BVC-PLL is in general more than one order of magnitude faster than ParaPLL (even for non-SIMD version)!

\section{Related Work}\label{sec:related}


\noindent{\bf Online and Parallel Shortest Path Distance Computation:}
The standard (single source) shortest path computation method is Dijkstra's algorithm~\cite{Dijkstra59} for weighted graphs and Breadth-First Search (BFS) traversal for unweighted graphs. 
There have been quite a list of efforts in designing parallel Dijkstra and BFS algorithms~\cite{Meyer:1998:DPS:647908.740136,LiuH7832842,Leiserson:2010:WPB:1810479.1810534}. Particularly,  
certain latest studies focus on performing multi-source or concurrent BFS over modern multi-core or GPU architectures~\cite{then2014more,liu2016ibfs}. However, it remains challenging to  answer the shortest path distance using these approaches due to the large traversal space for large graphs. 

\noindent{\bf Shortest Path Computation on Road Networks:}
Computing shortest path on road networks has been widely studied~\cite{Jing98,Jung02,Shekhar97,Sanders05,BastFSS07,Gutman04,GoldbergSODA05,HananSA08,Sankaranarayanan:2009,Kriegel:2008:HGE,Geisberger:2008:CHF,Tao11,Bauer:2010:CHG,Abraham:2010:HDS,Abraham:2011:HLA,Abraham:2012:HHL:2404160.2404164,Akiba:2014:FSD:2790174.2790188,Ouyang:2018:HME:3183713.3196913}, and has been applied successfully in industry practice. A more detailed review on this topic can be found in a recent survey~\cite{Bast2016}.
We note that the effectiveness of these approaches rely on the essential properties of road networks, such as the 
ones that are almost planar, have low vertex degree, are weighted, are spatial, or have a hierarchical structure, and they may not apply on scale-free complex networks, such as social and web graphs ~\cite{Gubichev:2010:FAE:1871437.1871503,Ouyang:2018:HME:3183713.3196913}. 


\noindent{\bf Theoretical Distance Labeling and Hop-based Labeling:}
There have also been several studies on estimating the distance between any vertices in large (social) networks~\cite{Potamias09,SarmaSMR10,Gubichev:2010:FAE:1871437.1871503, citeulike:7372248,zhao2011,DBLP:conf/icde/Qiao12}.
These methods fall within the group of distance-labeling~\cite{DBLP:journals/jal/GavoillePPR04},  
where the goal is to assign each vertex $u$ a label (for instance, a set of vertices and the distances from $u$ to each of them) and then estimate the shortest path distance between two vertices using the assigned labels.


The pioneering $2$-hop labeling method by Cohen {\em et al.}~\cite{cohen2hop} provides exact distance labeling on directed graphs.  However, numerous efforts over a ten-year period   ~\cite{hopiedbt,ChengYLWY06,DBLP:conf/sigmod/JinRXL12,DBLP:conf/edbt/ChengYLWY08,Abraham:2011:HLA,Geisberger:2008,Sankaranarayanan:2009} have largely failed in making $2$-hop labeling practical on large real-world graphs until the discovery of {\em Pruned Landmark Labeling (PLL)}~\cite{akiba2013fast}. 
In the past few years, a number of studies ~\cite{delling2014robust,li2017experimental} have further validated and confirmed the scalability of this approach. The idea has also been extended  to road networks~\cite{Akiba:2014:FSD:2790174.2790188} and out-of-core graph labeling~\cite{Jiang:2014:HDL:2732977.2732993}. 
Another direction of research involves the use of tree decomposition for shortest path distance computation ~\cite{DBLP:conf/sigmod/Wei10}, and particularly in utilizing it for hop-based labeling~\cite{Xiang:2014:AED:2673202.2673268,Chang2012,Ouyang:2018:HME:3183713.3196913}. 
There are also efforts that relax the distance computation to focus on cases when the distance is smaller than a certain threshold (useful for querying social networks)~\cite{DBLP:journals/pvldb/ChengSCWY12,DBLP:journals/corr/abs-1305-0507}.   

\noindent{\bf Others:} For related work on vertex-centric computation, please refer to Subsection~\ref{subsection:VC}. For recent progress on general parallel (VC-type) graph algorithms on modern computing architecture, please refer to ~\cite{Dhulipala:2018:TEP:3210377.3210414,Salihoglu:2014:OGA:2732286.2732294}. 
\vspace*{-3.0ex}
\section{Conclusion}\label{sec:conclusion}
In this paper, we proposed VC-PLL, which, to the best of our knowledge,  
is the first scalable parallelization of Pruned Landmark Labeling (PLL) that is able to produce the 
same result as the sequential method.  We have achieved this by mapping the algorithm to a 
vertex-centric model.  We also introduced a new batched execution mechanism for VC-PLL to better support message filtering and remote memory access. Based on the new model, we designed the BVC-PLL algorithm, which surprisingly can run faster than the original PLL as a sequential algorithm (demonstrated through both theoretical analysis and experimental validation).   
As far as we can tell, this is the first VC graph algorithm that can inherently run faster than its original counterpart even without parallelism.  Our experimental results further demonstrate the parallel efficiency and scalability of BVC-PLL and shows its superiority over the most recent Para-PLL algorithms on weighted graphs (using a straightforward extension of BVC-PLL).
In our future work, we plan to further investigate how to optimize BVC-PLL on weighted graphs and how to extend it for out-of-core graphs. We also plan to investigate the possibility of implementing the cost-saving mechanism in BVC-PLL for other graph algorithms.
\appendix
{\small


\noindent{\bf Lemma~\ref{lemma:messagefiltring}}
\bproof
To see this, we first need to prove that the shortest path distance $d(u,a)$ is smaller than or equal to  $d(u,v)+1$, where $a$ and $v$ is the direct neighbor of one another. By way of contradiction, let us assume $d(u,a) \geq d(u,v)+2$. Then, let $w$ be the highest rank vertex in $P_{ua}$, then, we can find a path from $u$ to $w$ to $v$ to $a$, which is $d(u,v)+1$. This suggests $d(u,a) \leq d(u,v)+1$. 
Next, we show $u$ indeed can reach $v$ in two consecutive iterations. 
This happens when $u$ reach $v$ via $a$ being the shortest path between $u$ to $v$: $d(u,v)=d(u,a)+1$; but $u$ is not the highest rank one in $P_{uv}$. Thus, $u$ is not added to $L(v)$ in $d(u,v)$ iteration. Now, assume $u$ reach $a^\prime$ with $d(u,a^\prime)=d(u,v)$ and $u \in L(v^\prime)$ (added in $d(u,v)$ iteration. If  $a^\prime$ is the neighbor of $v$, then $u$ will be sent to $v$ in $d(u,v)+1$ iteration as well.  \eproof

\noindent{\bf Lemma~\ref{lemma:labelmessageseset}}
\bproof 
Let $reach(u)$ be the subset of vertices $u$ reaches. In PLL, it corresponds to all the $(u, d(u,v))$ messages added into the $Q$ (Line $7$ in Algorithm~\ref{alg:DLD}). 
In VC-PLL, it corresponds to all the $(u,d(u,v))$ messages being sent to vertex $v$ (Line $5$ in Algorithm~\ref{alg:VertexCentricPLL_push}). 
Thus, $\bigcup_{u \in V} \{u\} \times reach (u)$ is the set consisting of all pairs $(u,v)$ for distance check. 
In PLL and VC-PLL, for a vertex $u$, it is assigned to the same subset of vertices (Corollary ~\ref{col:labelcoroll}). Also, it will also be sent to the same set of vertices which do not use $u$ as label. Thus, the set $\bigcup_{u \in V} \{u\} \times reach (u)$  is the same for both.  
\eproof

\noindent{\bf Theorem~\ref{thm:negativedistancecheckcost}}
\bproof
To quantify the difference of the time complexities between two algorithms, we focus on the cases where one algorithm can save computational cost when the $L^i(v)$ will be different for distance check $d(u,v)$. 

For the first case, let us consider vertex $x$, it has a vertex $u \in L^i(x)$. Now, consider any vertex $v \in B_i$ reaches vertex $x$ for distance check and returns negative result. If $v$ can reach $x$, it must be a label of neighbor $y$ of $x$, i.e., $v \in L^i(y), y \in N(x)$, and $v \notin L^i(x)$ (false distance check). When $v$ reaches $x$, it has also lower rank than $u$ but higher than $x$: $\pi(u)<\pi(v)<\pi(x)$. 
Given this, for PLL, $u$ is already in $L(x)$; however, for BVC-PLL, $v$ can reach $x$ before $u$ reaches $x$. Thus, this case will introduce a gain for BVC-PLL; and such $v$ is characterized and recorded in set $<x,u>$.

For the second case, let us consider vertex $y$, and it has a vertex $v \in L^i(y)$. Now, consider any vertex $u \in B_i$ reaches vertex $y$ for distance check and returns negative result. If $u$ can reach $y$, it must be a label of neighbor $x$ of $y$, i.e., $u \in L^i(x), x \in N(y)$, and $u \notin L^i(y)$ (false distance check). When $u$ reaches $y$, it has also higher rank than $v$: $\pi(u)<\pi(v)$. 
Given this, for BVC-PLL, $v$ is already in $L(y)$; however, for PLL, $u$ can reach $x$ before $v$ is added into $L(y)$. Thus, this case will introduce a gain for PLL; and such $u$ is characterized and recorded in set $<y,v>$. 
\eproof
}

\balance


\bibliographystyle{abbrv}
\bibliography{bib/ref,bib/graphdb,bib/reachability,bib/reachabilitytods,bib/ComplexNetwork,bib/3hop,bib/Proposal,bib/socialnetwork,bib/distance,bib/haixun,bib/simplification,bib/ref-spx,bib/Yangdissertation}  








\end{document}